\documentclass[article]{elsarticle}

\usepackage{lineno,hyperref}
 \usepackage{booktabs} 
 \usepackage{array}
 \usepackage[T1]{fontenc}
\newcolumntype{H}{>{\setbox0=\hbox\bgroup}c<{\egroup}@{}}

\journal{Journal of \LaTeX\ Templates}

\begin{document}

\begin{frontmatter}

\title{Rethinking  travel behavior modeling representations through embeddings}
\tnotetext[mytitlenote]{Fully documented templates are available in the elsarticle package on \href{http://www.ctan.org/tex-archive/macros/latex/contrib/elsarticle}{CTAN}.}

\author{Francisco C. Pereira}
\address{Technical University of Denmark\\Bygningstorvet 116B-123A , Kongens-Lyngby, Denmark\\camara@dtu.dk}



%\address[mymainaddress]{}

\begin{abstract}

%
%story line:
%- Explanation of embeddings
%- Baseline models (dummies and PCA)
%- Embedding cases
%- Total model - so so (?)
%	- quick discussion - more stability (show 2015 and 2016 tables)
%	- better generalization (?) 
%- Advantages of embeddings
%- dataset size
%- dataset skewness 
%- interpretability
%

This paper introduces the concept of travel behavior embeddings, a method for re-representing discrete variables that are typically used in travel demand modeling, such as mode, trip purpose, education level, family type or occupation. This re-representation process essentially maps those variables into a latent space called the \emph{embedding space}. The benefit of this is that such spaces allow for richer nuances than the typical transformations used in categorical variables (e.g. dummy encoding, contrasted encoding, principal components analysis). While the usage of latent variable representations is not new per se in travel demand modeling, the idea presented here brings several innovations: it is an entirely data driven algorithm; it is informative and consistent, since the latent space can be visualized and interpreted based on distances between different categories; it preserves interpretability of coefficients, despite being based on Neural Network principles; and it is transferrable, in that embeddings learned from one dataset can be reused for other ones, as long as travel behavior keeps consistent between the datasets. 

The idea is strongly inspired on natural language processing techniques, namely the word2vec algorithm. Such algorithm is behind recent developments such as in automatic translation or next word prediction. 

Our method is demonstrated using a model choice model, and shows improvements of up to 60\% with respect to initial likelihood, and up to 20\% with respect to likelihood of the corresponding \emph{traditional model} (i.e. using dummy variables) in out-of-sample evaluation. We provide a new Python package, called PyTre (PYthon TRavel Embeddings)\footnote{https://github.com/camaraf/PyTre}, that others can straightforwardly use to replicate our results or improve their own models. Our experiments are themselves based on an open dataset (swissmetro \cite{bierlaire2001acceptance}).

\end{abstract}

\begin{keyword}
\texttt{text embeddings}, \texttt{travel behavior}, \texttt{machine learning}, \texttt{latent representations},
\end{keyword}

\end{frontmatter}

%\linenumbers

\section{Introduction}

Since their early days, representation in random utility behavior models has followed generally quite clear principles. For example, numeric quantities like travel time and cost may be directly used or transformed depending on observed non-linear efects (e.g. using log). Numeric variables that are not ``quantities" per se, such as age or even geographic coordinates tend to be discretized and then transformed into vectors of dummy variables. Similarly, categorical variables such as education level or trip purpose are already discrete, and thus are also usually ``dummyfied". Then, we may interact any subset of the above by combining (typically, multiplying) them, as long as we get in the end a vector of numeric values that can be incorporated in a statistical model, a linear one in the case of the most common logit model. 

There are however phenomena that are hard to represent, and modelers end up struggling to find the right representation. For example, influence of social interactions between different persons, hierarchical decision making, autocorrelated nature of time and space, or abstract concepts such as accessibility, attitudes, personality traits and so on. The point here, is that the nature of our models seems to enforce a compromise between the true semantics of a variable (i.e. the ``meaning" of a certain information for the decision making process) and its realisation in practice. And that further research should be done to find new representation paradigms.  

Historically speaking, the natural language processing (NLP) field has had similar dilemmas for decades, and for a while two general trends were competing: the statistical modeling approaches, and the linguistic theory based approaches. The former relied on simple representations, such as vector frequencies, or dummy variables, to become practical, while the latter used domain knowledge such as grammars or logic. Until recently, neither had considerable success in making machines able to understand or generate human language\footnote{Although interestingly they pushed forward other areas, such as computer programming language design, or automated model verification.}, but developments in deep neural networks together with overwhelmingly massive amounts of data (i.e. the World Wide Web) brought them to a new area, where the two are approaching each other and achieving hitherto results considered extremely hard, such as question answering, translation, next word prediction. One of the key concepts in this revolution is that of embeddings, which will be further explained in this paper. 

Our focus here is on the representation of categorical variables. The default paradigm is dummy variables (also known as ``one-hot-encoding" in machine learning literature), which have well-known limitations, namely the explosion of dimensionality and enforced ortogonality. The former happens because we assign one new ``dummy" variable to each of D-1 categories, and easily go from a small original variable specification to one with hundreds of variables, bringing problems in model estimation and analysis. This often affects the data collection process itself. Since one doesn't want to end up with too many categories, we might as well give less options in a survey, or decrease the resolution of a sensor. The problem of enforced ortogonality relates to the fact that, in a dummy encoding, all categories become equidistant. The similarity between ``student" and ``employed" is the same as between ``student" and ``retired", which in many cases (e.g. mode choice, departure time choice) goes against intuition. Other encoding methods exist, such as contrasted encoding or principal components analysis (PCA). The former ends up being a subtle variation on the dummy approach, but the latter already provides an interesting answer to the problem: categories are no longer forcibly equidistant, and the number of variables can be much reduced. However, it is a non-supervised approach. The distance between ``student" and ``employed" will always be the same, regardless of the problem we are solving, but this may be intuitively illogical if we consider car ownership versus departure time choice models for example. 

The key idea in this paper is to introduce a method, called Travel Behavior embeddings, that borrows much from the NLP concept. This method serves to encode categorical variables, and is dependent on the problem at hand. We will focus on mode choice, and test on a well-known dataset, by comparing with both dummy and PCA encoding. All the dataset and code are made openly available, and the reader can follow and generate results\footnote{We have to forewarn the reader that the embeddings training method is itself stochastic, so the reader will necessarily generate results that are similar to ours within an $\epsilon$ difference.} him/herself using an iPython notebook included. Our ultimate goal is certainly that the reader reuses our PyTre package for own purposes. 

This paper presents some results and conclusions, after a relatively long exploration and analysis process, including other datasets and code variations not mentioned here for interest of clarity and replicability. While we show these concepts to be promising and innovative in this paper, one should be wary of over-hyping yet another Machine Learning/Artificial Intelligence concept: after all, Machine Learning is still essentially based on statistics. In NLP, the number of different words in consideration at a given moment can be in order of tens of thousands, while our categorical variables rarely go beyond a few dozens. This means that, for example, it becomes clear later that the least number of original categories, the less the benefit of embeddings (in the limit, a binary variable like gender, is useless to do embeddings with), and also that if we do get a significantly large and statistically representative dataset, a dummy variables representation is sufficient. We will quickly see, however, that complexity can grow quick enough to justify an embeddings based method even if without the shockingly better performance observed in NLP applications.

\section{Representing categorical variables}

We are generally concerned with random utility maximization (RUM) models, for they have a dominant role in travel behavior modeling. The nature of such models is predominantly numeric, linear, and quite often strictly flat (notwithstanding hierarchical variations, such as nested models \cite{ben1985discrete}, hierarchical Bayes \cite{arora1998hierarchical}, or non-linear transformations). As a consequence, while numerical variables (e.g. travel time, cost, or income) can be directly used as available, perhaps subject to transformations or segmentation, nominal ones bring about a greater challenge. We tend to enforce a limited set of treatments such as:
\begin{itemize}
\item Dummy variables, or one-hot encoding - for each categorical variable $v$ with D categories, we get D-1 binary variables (the ``dummies"). At each input vector $x_n$, with categorical value $v=d$, the value ``1" will be assigned to the corresponding dummy, while ``0" to all others. If $v$ corresponds to the ``default" category, all dummies are ``0".
\item Contrast encoding \cite{davis2010contrast} - same as dummy encoding, but instead of ``1" for each category, we have a value that results from a \emph{contrasting} formula. There are many different formulas (e.g. Helmert, Sum, Backward Difference), but all consist of subtracting the mean of the target variable, for a given category, with a general stastic (e.g. the mean of the dependent variable for all categories; the mean of the dependent variable in the previous category in an ordered list).
\item Principal Components Analysis (PCA) - run the PCA algorithm on the data matrix obtained by dummy representation of the categorical variable, then re-represent it with the corresponding projected eigenvector coefficients. One selects K eigenvectors (e.g. according to a variance explained rule), and thus each category is mapped to a vector of K real values. 
\item Segmenting models, mixture models - A general alternative to categorical data representation is in fact to avoid it in the first place. One obvious method would be through creating hierarchical disaggregate methods (e.g. one per category). This is not in itself a representation paradigm, but an alternative way to see this problem. It certainly raises scalability and inference concerns. 
\end{itemize}

In datasets where behavior heterogeneity is high, and number of observations is significantly smaller than population size, increasing dimensionality by adding a variable per each category is very risky because the amount of data that is in practice usable to estimate each new coefficient becomes insufficient. A simple intuition here is by considering that, for a dummy variable that is only ``1" for a few observations in the dataset, its coefficient will be ``activated" only that small number of times. If there is a lot of variance in the associated behavior, the variance of the coefficient will also be large, and the coefficient will be considered statistically insignificant.

The benefit of representations that map into a latent space, like embeddings and PCA, is that such a space is inevitably shared, and thus every observation contributes indirectly to all category variables. This comes with no interpretability cost, because one can always map to the ``dummy" space and analyse the individual coefficients, as will be shown in our experiments. 

%As a consequence of this compartmentalization, the semantic richness associated to such data is regarded to a very little extent. For example, the concept of \emph{family with children} is associated to a certain collection of travel behaviors (e.g. need for more luggage space; pick up/drop off trips; tighter constraints on departure and arrival times) that demand careful consideration. As a starting point, we could take advantage of semantic similarity between different categories in a high dimensional space, where, for example, a \emph{family with children} is closer to a \emph{single person with children} than to a \emph{single person}. If not more, such re-representation should help create semantically richer models. 

\section{The concept of text embeddings}\label{concept}

The idea of text embeddings comes from a simple re-representation necessity. A natural-language processing system is itself also a numeric machine, therefore it requires each individual word in a dictionary to match its own numeric representation. Just as in our travel models, a possible solution has been to use dummy variables, and it is quite obvious that the dimensionality of such a one-hot encoding vector, quickly becomes overwhelming. Think for example next word prediction algorithm, like the one we have in our smartphones. It is essentially a skip-gram \cite{guthrie2006closer} model that predicts the next word, given the n words before. The English dictionary has about 300000 words, and if we have about 5 words before for context, the number of independent variables of the model would become 1.5 million! 

The goal of text embeddings algorithms (e.g. Word2Vec \cite{mikolov2013efficient}) is to a) reduce the representation of each word to a computationally acceptable dimension, while simultaneously b) learning the semantic distance between different words. In other words, the euclidean distance of semantically related words (e.g. ``dog" and ``cat") in this new space should be smaller than unrelated words (e.g. ``dog" and ``optimize"). As mentioned before, in a dummy (or one-hot) encoding, all distances between words are equal by definition. 

The word embeddings methodology is very well explained in several webpages such as \cite{mccormick_2016}, so the reader is strongly encouraged to visit them first. However, for the sake of completeness, we summarize here the general idea. 

\begin{figure}[htbp]
\begin{center}
\includegraphics[width=4in]{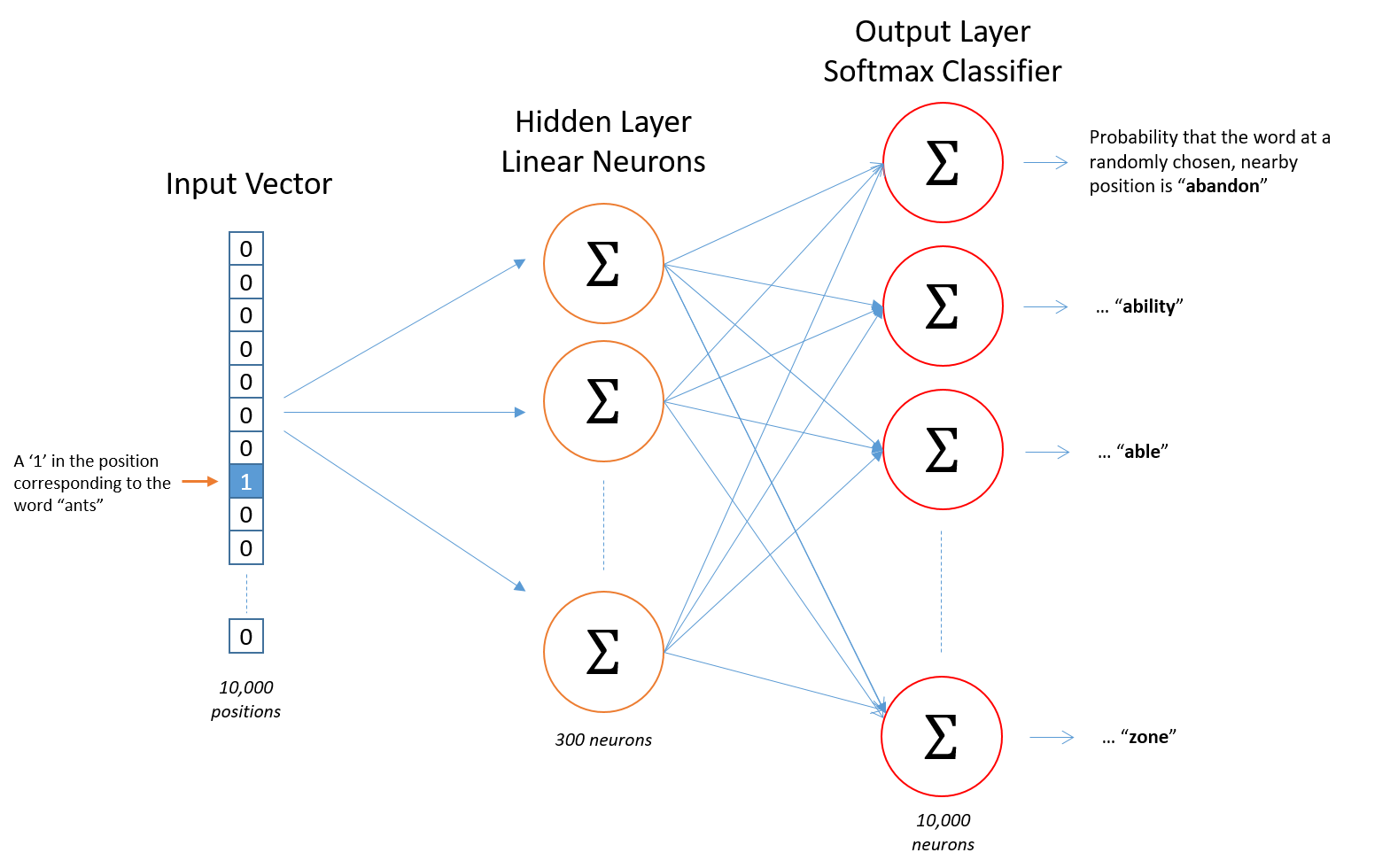}
\end{center}
\caption{The skip gram architecture \cite{mccormick_2016}}\label{skip_gram}
\end{figure}

Imagine the following task: given a word $w_i$ in a text, predict the next word $w_o$. If we solve it with a neural network model, we could have the architecture in Figure \ref{skip_gram}, where the input consists simply of the one-hot-encoding representation of the word (i.e. one dummy variable for each word in a dictionary of dimensionality $D$), and the output corresponds to the probability of each word in the dictionary being the next one (also a vector with dimensionality $D$).

The output layer thus consists simply of a softmax function. In other words, exactly the \emph{classical} multinomial logit formulation that we would have in an RUM, in which each different word corresponds to an ``alternative". 

The concept of embeddings is directly associated to the hidden layer, which is a set of linear activation neurons, typically with a dimensionality $K<<D$. Each such neuron is simply an identity function: it sums all inputs; then propagates this sum to the output layer. Since only \textbf{one} input neuron is activated at a time (remember that the input is a one-hot-encoding vector, with one ``1" and the rest with ``0"), each hidden layer neuron just propagates the (single) weight that links to that input neuron. If we have enough data for training this model, we will eventually land on a situation where, for each input word, there is a fixed vector of weights that are directly used in the output (softmax) function, to generate the prediction. With more data, this weight vector will not change (down to some small delta threshold). These stable vectors are what we call \textbf{embeddings}, and the dimensionality of these vectors is called \textbf{embedding size}.

Formally, we have a dataset $\mathcal{D}=\{x_n, y_n\}, n=1\ldots N$, where each $x_n$ and $y_n$ are one-hot (dummy) encodings of categorical variables\footnote{When $x_n=y_n$, we call this an \emph{autoencoder}.}. The dimensionality of $x_n$ is $D\times1$, with $D$ being the number of different categories in $x_n$, while  the dimensionality of $y_n$ is $C\times1$, with $C$ being the number of categories (alternatives) in $y_n$. The full expression for the embeddings model as described is: 

\[p(y_n=c|x_n)=\frac{e^{B_cWx_n+\alpha_c}}{\sum_{j=1}^C e^{B_jWx_n+\alpha_j}}\]

where $W$ is the embeddings matrix of size $K\times D$, where $K$ is called the \emph{embeddings size}. $B$ is a matrix of coefficients ($C\times K$) for the softmax layer, so $B_c$ is simply the coefficients (row) vector for output class (alternative) $c$, and $\alpha_c$ is the corresponding intercept. The typical loss function used in such models is called the \emph{categorical cross entropy}:
\[\mathcal{L}(n)=-\sum_{c=1}^C \delta_{\{y_n=c\}}\log p(y_n=c|x_n)\]

Where $\delta_{i}$ is the \emph{kronecker delta} ($\delta_{true}=1; \delta_{false}=0$), and $\mathcal{L}(n)$ is the cumulative loss for an individual data point. This formalization is the simplest version, without loss of generality. In practice, as seen below, we will model multiple embeddings matrices simultaneously, and will add regularization terms to the loss function, so the models tested in this  paper consist of compositions of the above. 

%input vector
%linear layer
%softmax
%cross entropy objective function

\begin{figure}[htbp]
\begin{center}
\includegraphics[width=4in]{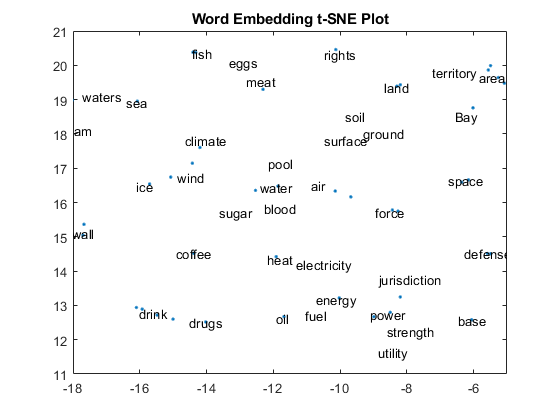}
\end{center}
\caption{Visualization of a subset of words from FastText word embeddings database \cite{fasttext_2019}}\label{fastext_2019}\label{wordembs}
\end{figure}

So these so called \emph{embeddings} are in fact a relatively shallow data representation in a simple neural network. What is their added value? Obviously, the first practical benefit is dimensionality reduction, because now there is a mapping between each of the $C$ words to a unique vector of size $K$. The second aspect is that this new representation is the one that maximizes the performance towards a specific task (in our example, prediction of the next word), therefore it is a supervised process, as opposed for example to PCA. The third and more interesting aspect relates with semantic similarity. A natural consequence of the mentioned algorithm is that words that have similar output distributions (i.e. next words) will tend to be close to each other. Figure \ref{wordembs} shows a 2D visualization (t-SNE) with a subset of english words. In such a visualization, data is projected in 2D space by maintaining the same vector-to-vector distances as in the original ($K$ order space). Therefore the X and Y axes have no specific meaning, only distances between every pair of points are relevant. 

\begin{figure}[htbp]
\begin{center}
\includegraphics[width=5in]{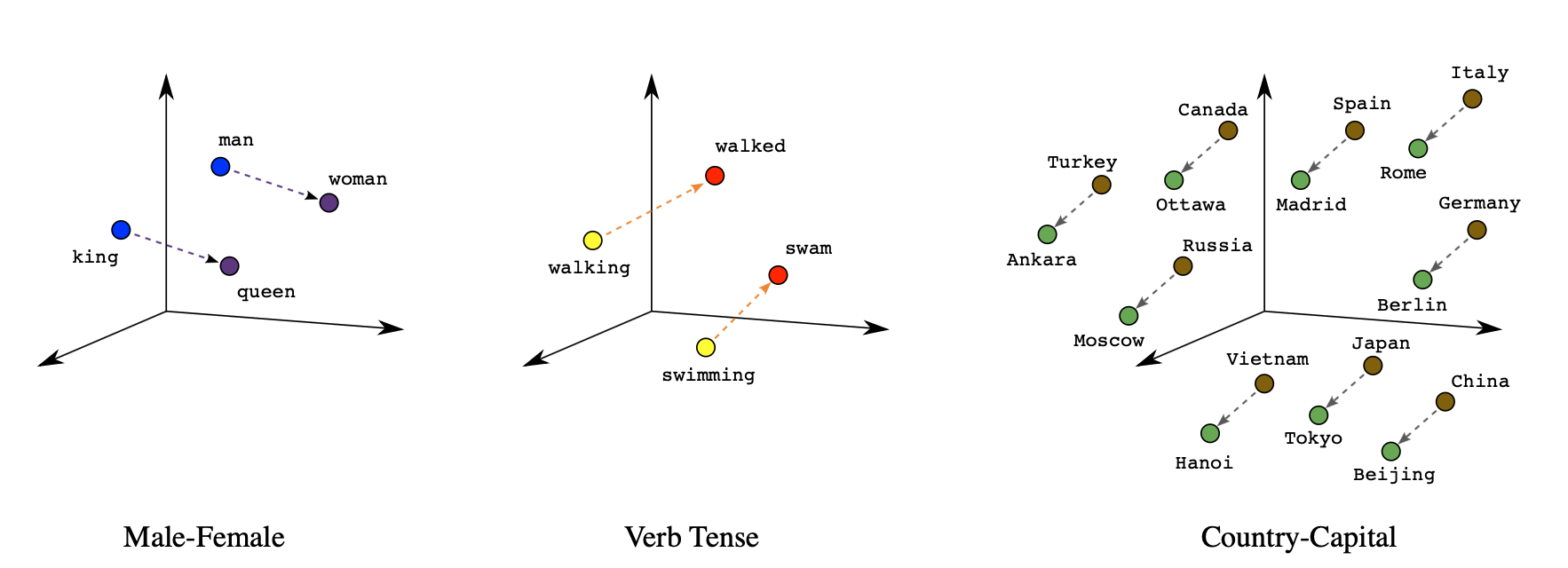}
\label{algebra}
\end{center}
\caption{Some classical examples of embeddings algebra \cite{translating_2019}}
\end{figure}

We can see that semantically similar concepts, more specifically concepts that tend to have the same distribution of ``next words", are placed closer. Another intriguing consequence is that, since the words are now in the $K$ dimensional, embeddings space, we can also do some linear algebra on them. A well known formulation is $King-Man+Woman=Queen$. Essentially, the vector $King-Man$ corresponds to the concept of ``crowning" (therefore $Woman+crowning=Queen$). The same could be done with many other concept pairs. Figure \ref{algebra} show also an alternative interpretation of ``man-female", as well as examples with cities and verb tense. 

Finally, another relevant note on the embeddings representation is that, just like the PCA encoding, one can always project back into the original space and use this for interpretability. In other words, since there is a 1-to-1 mapping from each category to its encoding, there is also a 1-to-1 mapping between a model that uses dummy variables and a model using such encodings. This may be useful for interpretability, since in the case of dummy variables we have a direct interpretation (e.g. a beta coefficient value in a logit model) for the effect of a given category, while the same doesn't happen for an encoded variable (i.e. there is no \emph{meaning} for the value of a single beta coefficient in an embeddings encoding when K>1). In order to preserve statistical significance information (e.g. \emph{p-values}) we only need to follow the well known rules of normal random variables\footnote{If $X\sim \mathcal{N}(\mu_X,  \sigma_X^2)$ and $Y\sim \mathcal{N}(\mu_Y, \sigma_Y^2)$, then $X+Y\sim \mathcal{N}(\mu_X+\mu_Y, \sigma_X^2+\sigma_Y^2)$ and $kX\sim \mathcal{N}(k\mu_X, k^2\sigma_X^2)$, for any constant $k$.}.

There are open databases available (e.g. GLoVe \cite{goldberg2014word2vec}, FastText \cite{fasttext_2019}) that provide word embedding tables for the entire English language (Glove provides several embedding tables, up to embedding size between 100 and 300). In our next word application example, we now talk about models with 500-1500 variables, which is very manageable for our machines today. 

Summarizing, the general idea of word embeddings is to re-represent a categorical variable into a lower dimensional representation with continuous values \footnote{The concept of \emph{embeddings} in Deep Learning literature is actually broader than what we discussed here. It applies to any internal stable representation of the input data. In other words, for the same input vector, there will be a (typically lower dimensional) fixed set of weights somewhere in one or more layers in the model.}. Whenever such a variable is to be used in a model, one can simply replace it with the corresponding embeddings vector. We have previously demonstrated the value of such word embeddings in demand prediction in special events \cite{rodrigues2019combining}, where we collected event textual descriptions, and used Glove embedding vectors to incorporate such information in a neural network model. 

Finally, an interesting point to mention relates to the typical difference in dataset size between the original embeddings training model (Glove, approximately 6 billion input word vectors from 37 million texts) and the model one implements to solve a particular problem (in our special events case, less than 1000 short event descriptions, with at most few hundred words each). Instead of creating ourselves a new embeddings model using the events dataset, we reused the pre-trained GloVe dataset.  The benefit is significant because in practice we trained our model to deal with all words in the dictionary, much beyond the limited vocabulary that we obtained in our 1000 short texts. In practice we have used a very small percentage of the english dictionary. When, in an out-of-sample test, our model finds words that were not in the training set, it still works perfectly well.

\section{Travel behaviour embeddings}

Differently to textual data, our goal in this paper is to explore the large amount of categorical data that is often collected in travel surveys. This includes trip purpose, education level, or family type. We also consider other variables that are not necessarily of categorical nature, but typically end up as dummy encoding, due to segmentation, such as age, income, or even origin/destination pair. 

Our hypothesis is that, given the limitations of dummy variables that are commonly used and the unsupervised nature of PCA, using instead an embeddings mechanism should improve significantly the quality of our models, both in terms of loglikelihood but also in terms of allowing for lower complexity (i.e. less variables). Ultimately, one could think of a framework such as GLoVe, where embeddings for such variables could be trivially shared with the community. For example, we could have a ``Travel behavior embeddings" database, incrementally built from travel surveys from around the world. Such database could have embeddings for mode choice target variables, but also for departure time, destination choice, car ownership, and so on. Whenever a modeler wanted to estimate a new model, she could just download the right encodings and use them directly. This is particularly relevant if one considers the complicated challenges for opening or sharing travel survey datasets in our field. Of course, a major question arises: are behaviors that consistent across the world? There are certainly nuances across the world, but we believe that general patterns would emerge (e.g. a ``business" trip purpose will be closer to ``work" than ``leisure", in a departure time choice model; ``student" will be closer to ``unemployed" than to ``retired" in a car ownership model).

\subsection{The general idea}

We believe that, as with word embeddings, a mapping that preserves semantic distance relative to a certain choice problem, should be useful for modeling. As with a PCA encoding, another benefit is that by sharing parameters in the learning process, the model can generalize better, as opposed to a dummy encoding, where each categorical value has its own parameter, that is only active when observed.  

\begin{figure}[htbp]
\begin{center}
\includegraphics[width=3in]{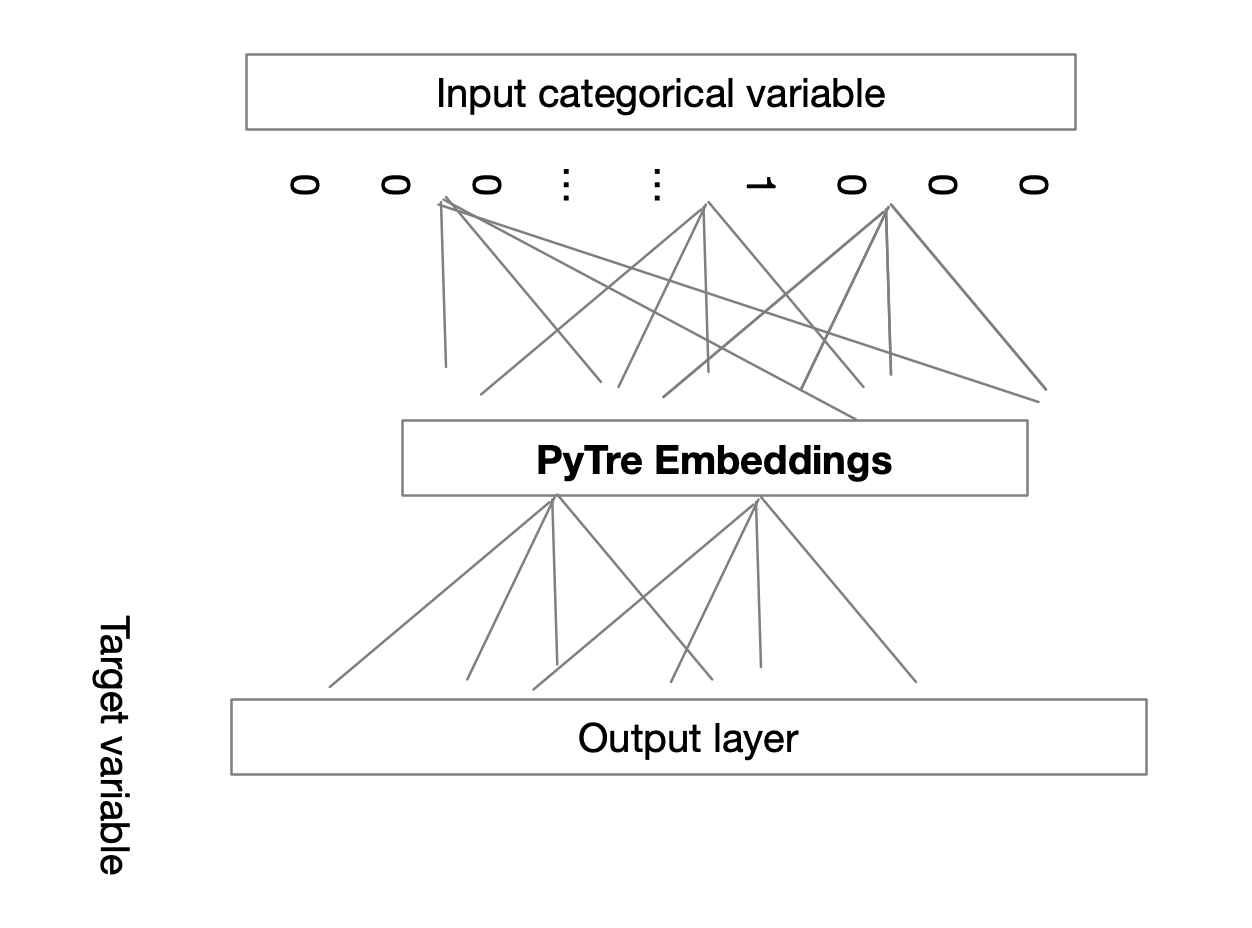}

\end{center}
\caption{The general idea}\label{generalidea}
\end{figure}

The general idea is thus to create a mapping between a variable for which we want to find an embeddings representation, and a target variable, as in Figure \ref{generalidea}. We call the mapping function ``PyTre Embeddings", because that is the name of the object in our proposed Python ``Travel Embeddings" package. 

%
%From an experimental design and application perspective, the typical approach should be the following:
%\begin{enumerate}
%\item For each variable of interest (categorical, or segmented numeric):
%\begin{enumerate}
%\item If there is an available embeddings table, 
%\begin{itemize}
%\item encode it in both training and test set by replacing each entry by its corresponding embeddings vector \footnote{function \emph{encode()} in PyTre package.}
%\end{itemize}
%\item else 
%\begin{itemize}
%\item add variable to list \emph{emb\_vars}
%\end{itemize}
%\end{enumerate}
%\item Split dataset into embeddings train set, choice model train set and test set
%\item For each variable in \emph{emb\_vars}:
%\begin{enumerate}
%\item Learn the new embeddings using the embeddings train set \footnote{function \emph{fit()} in PyTre package} %This should be done simultaneously (all variable embeddings estimated at once).
%\end{enumerate}
%\item Encode choice model train and test set \emph{emb\_vars} using the trained embeddings
%\item Estimate choice model accordingly using its train set
%\item Evaluate the new model using the test set
%\end{enumerate}

From an experimental design and application perspective, the approach followed in this paper is the following: 
\begin{enumerate}
\item Create list of categorical variables to encode  (the \emph{encoding set})
\item Split dataset into train, development and test sets
\item For each variable in \emph{encoding set}, learn the new embeddings using the embeddings train set \footnote{function \emph{fit()} in PyTre package}. This should be done simultaneously (all variable embeddings estimated at once, as explained in the next section).
\item Encode choice models for train, development and test sets using the learned embeddings
\item Estimate choice model accordingly using its train set
\item Evaluate the new model using the test set
\end{enumerate}

Since there is stochasticity in the embeddings training model, we will repeat the above multiple times, for the different experiments in the paper (and report the respective mean and standard deviation statistics). Whenever we want to analyse a particular model (e.g. to check the coefficients of a choice model), we select the one with the highest likelihood at the development set (i.e. in practice, its out-of-sample generalization performance), and report its performance on the test set.  

%This paper aims to be an example of such approach. We will now explain the Travel Embeddings methodology implemented in the package that we designed, called PyTre. 

%Dimensionality reduction is not as radical as in word embeddings

\subsection{Methodology}

\begin{figure}[htbp]
\begin{center}
\includegraphics[width=4in]{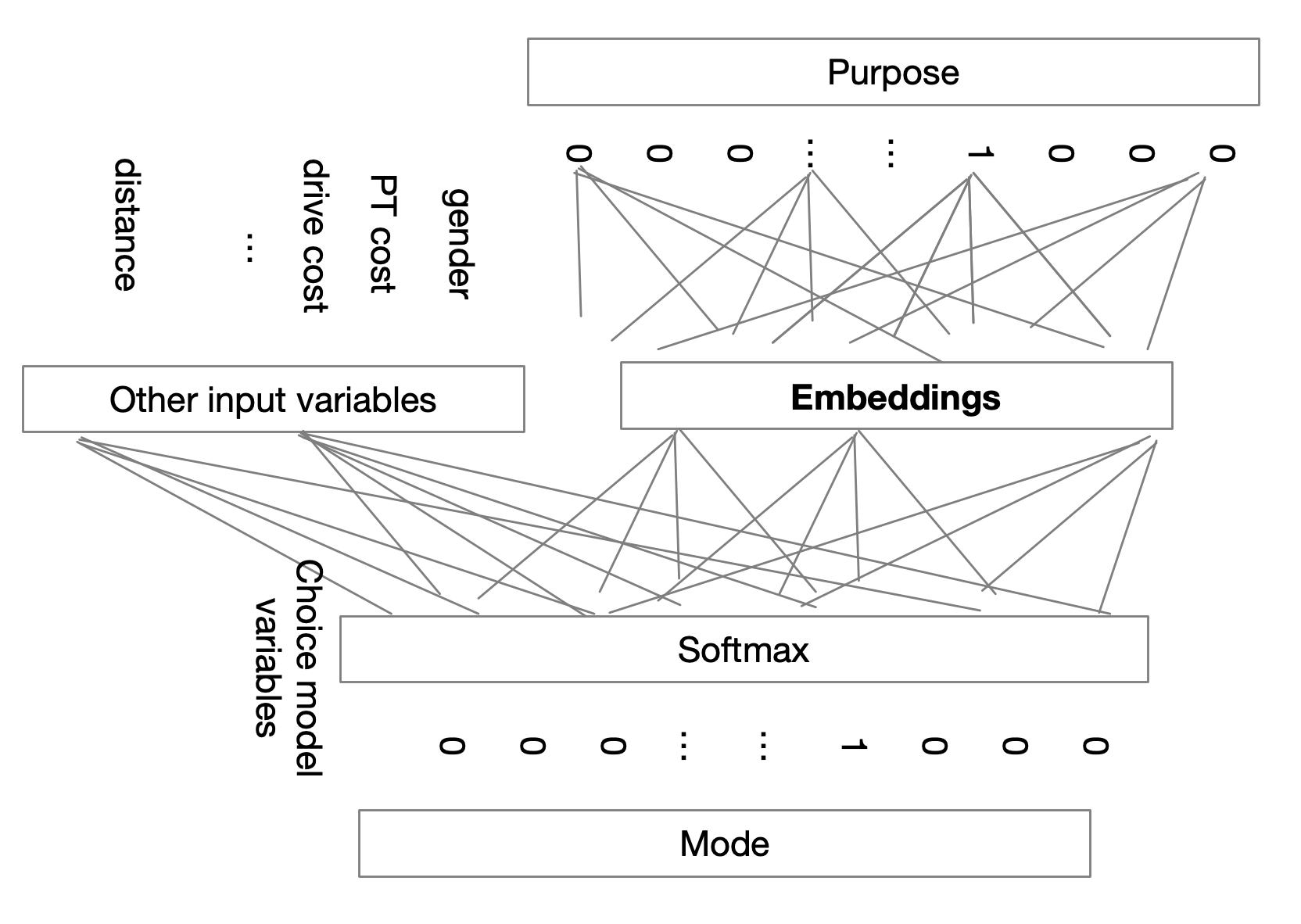}

\caption{Travel embeddings model}\label{TravelBehaviorEmbeddings_No_reg}
\end{center}
\end{figure}

Since a choice model will typically involve other variables than the categorical ones that we learn the embeddings for, it is important to take into account their effects. Figure \ref{TravelBehaviorEmbeddings_No_reg} shows the simplest travel embeddings model. As an example, the categorical variable is \emph{trip purpose}, and there are a few other variables such as gender, cost of the alternatives, distance, and so on. Notice that they are directly fed into the softmax output layer, together with the embeddings output. 

\begin{figure}[htbp]
\begin{center}
\includegraphics[width=2.35in]{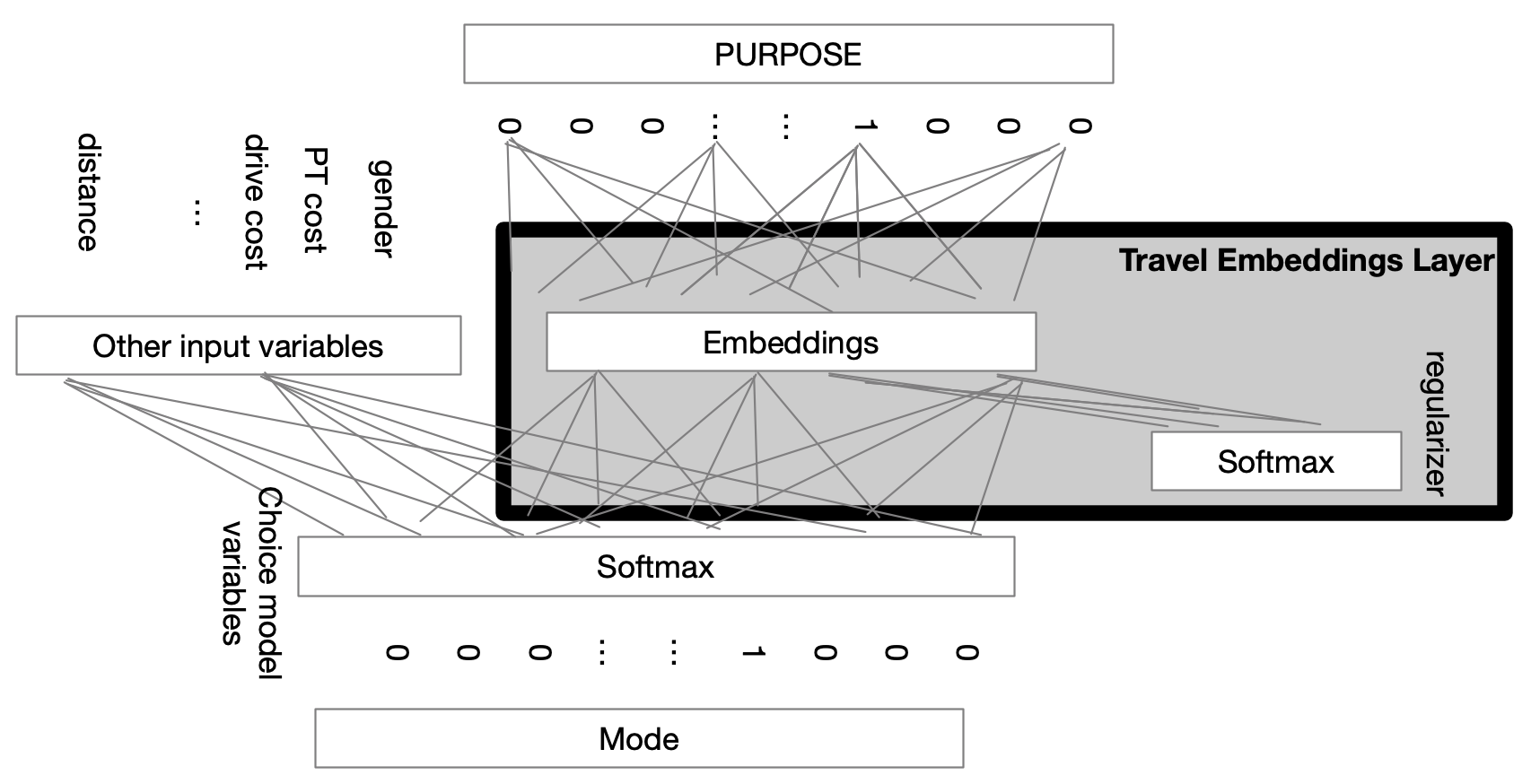}
\includegraphics[width=2.35in]{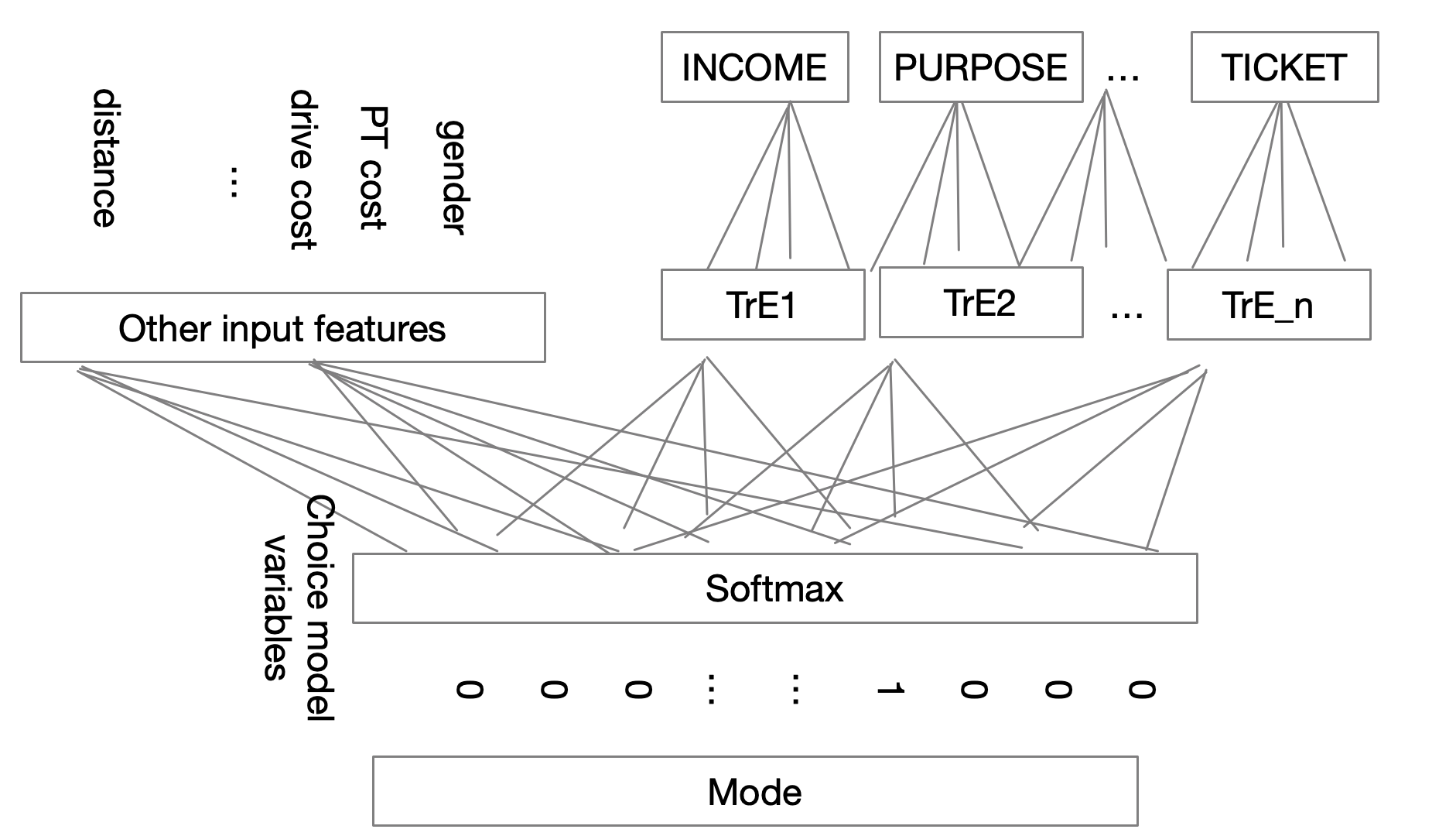}

\caption{Travel embeddings model with regularization (left); Complete model, combining multiple travel embeddings layers (right).}\label{TravelBehaviorEmbeddings}
\end{center}
\end{figure}

The dataset sizes in transportation behavior modeling are substantially smaller than typical word embeddings ones, and the risk of overfitting is therefore higher. To mitigate this problem, besides adding regularization penalties in the objective function, we add what we call a \emph{regularizer} layer for each embedding, which is no more than a softmax layer that penalizes whenever it cannot recover the original one-hot-encoding vectors (Figure \ref{TravelBehaviorEmbeddings}, left). We call the combination of embeddings and its regularizer network, a \emph{Travel Embeddings layer}. Finally, it is obviously better to train all embeddings simultaneously, so that they accommodate each other's effects (Figure \ref{TravelBehaviorEmbeddings}, right). 

%Automatically learning embedding size
%
%Some rules for selecting embedding size (1:N-1, otherwise in N equals simply averaging); grouping is good; 

%\begin{figure}[htbp]
%\begin{center}
%\includegraphics[width=2in]{TravelBehaviorEmbeddings.png}
%\label{}
%\end{center}
%\end{figure}

\section{An experiment with mode choice}

The goal of this paper is to test the potential of embeddings in a simple and well-known choice model context, comparing it to well-known baseline techniques. Therefore, the general model specification follows quite simple assumptions. We expect that in future work from us or others, more elaborate derivations can take advantage of embeddings such as nested, mixed logit or latent class choice models (LCCM), for example. 

We will apply the methodology to the well-known ``Swissmetro" dataset. We will compare it with a dummy variables and PCA baselines. We will follow the 3-way experimental design mentioned before: split the dataset into train, development and test sets, so that the embeddings, PCA eigenvectors and the choice model are estimated from the same train and development sets, and validate it out-of-sample. For the sake of interpretability, we will also project back coefficients from the embeddings as well as PCA models into the dummy variable space. 

All experiment code is available as a \emph{jupyter notebook} in a package we created for this work (to which we called PyTre). For estimating the multinomial logit model (MNL) we used the PyLogit \cite{brathwaite2018asymmetric} package.

\subsection{The Swissmetro dataset}

The Swissmetro dataset consists of survey data collected on the trains between St. Gallen and Geneva, Switzerland, during March 1998. According to its description \cite{bierlaire2001acceptance}, the respondents provided information in order to analyze the impact of the modal innovation in transportation, represented by the Swissmetro, a revolutionary mag-lev underground system, against the usual transport modes represented by car and train. After discarding respondents for which some variables were not available (e.g. age, purpose), a total of 10469 responses from 1188 individuals were used for the experiments.

We split the dataset into 3 different parts:
\begin{itemize}
\item Embeddings train set: 60\% of the dataset (6373 vectors)
\item Development set: 20\% of the dataset (2003 vectors)
\item Test set: 20\% of the dataset (2003 vectors)
%\item Dataset A: From 2006 to 2014, giving a total of 1047336 trips. This dataset serves \textbf{exclusively} to learn the embeddings, for each categorical variable. 
%\item Dataset B: Year of 2015, giving a total of 165984 trips. This dataset serves to do embeddings model selection. In other words, we train the embeddings models on Dataset A, and evaluate them  in Dataset B. The chosen embeddings are those that perform best in Dataset B. In this way, we we look for the best out-of-sample generalization capability. 
%\item Dataset C: A random subset of the year of 2016, giving a total of 25998 trips\footnote{ This number is uniquely determined by memory limitations. We use PyLogit \cite{brathwaite2018asymmetric}.}. This dataset is the one used to both estimate the baseline mode choice model, and the embeddings one.
%\item Dataset D: Second random subset of 2016, giving a total of 11142 trips. This dataset is used to compare the out-of-sample performance of the embeddings \emph{versus} baseline model. There is obviously no shared data between datasets C and D. 
\end{itemize}

%We shuffle the dataset but keep  order since each respondent replies multiple times (that are contiguous in the dataset) in order not to have the same respondent in different splits. 

\subsection{Principles for the model specification }
The PyLogit package \cite{brathwaite2018asymmetric} also uses Swissmetro as an example. Therefore, our model specifications will extend the default one from this package. We re-estimated this model with the train set and validated with testset. The results are shown in tables \ref{original_1} and \ref{original_2}. Since we are comparing the models at the test set, the key indicators should be pseudo R-square and log-likelihood. Indicators that consider model complexity (robust r-square and AIC) are less important on the test set in our view because the overfitting effect (i.e. improving fit just by adding more variables) will no longer be verifiable in this way. Instead, one sees overfitting if test set performance is considerably inferior to the training set.

\begin{center}

\begin{table}[!htbp]
\scriptsize
\begin{tabular}{lclc}
\toprule
Dep. Variable:                                                 &          CHOICE         & No. Observations: &   6,373     \\
Model:                                                         & MNL & Df Residuals:    &   6,359     \\
Method:                                                        &           MLE           & Df Model:        &     14      \\
Date:                                                          &     28 Aug 2019    & Pseudo R-squ.:   &   0.284     \\
Time:                                                         &         23:01:01        & Pseudo R-bar-squ.:  &   0.282     \\
AIC:                                                           &        9,419.631        & Log-Likelihood:  & -4,695.816  \\
\textbf{Pseudo R-squ. (testset):} &   \textbf{0.279}      & \textbf{ Pseudo R-bar-squ. (testset):} &   \textbf{0.274}     \\
\textbf{AIC (testset):}                                                           &        \textbf{3,000.5}        & Log-Likelihood (testset):  & \textbf{-1,486.2} \\

\bottomrule
\end{tabular} \caption{Multinomial Logit Model Regression Results - original model}\label{original_1}
\end{table}
\begin{table}[!htbp]
\scriptsize
\begin{tabular}{lccHcHH}
                                                                        & \textbf{coef} & \textbf{std err} & \textbf{z} & \textbf{P$>$$|$z$|$} & \textbf{[0.025} & \textbf{0.975]}  \\
\midrule
ASC Train                                                      &      -0.5660  &        0.158     &    -3.575  &         0.000**        &       -0.876    &       -0.256     \\
ASC Swissmetro                                                 &      -0.3532  &        0.121     &    -2.929  &         0.003**        &       -0.590    &       -0.117     \\
Travel Time, units:hrs (Train and Swissmetro)                  &      -0.6815  &        0.041     &   -16.432  &         0.000**        &       -0.763    &       -0.600     \\
Travel Time, units:hrs (Car)                                   &      -0.7186  &        0.049     &   -14.752  &         0.000**        &       -0.814    &       -0.623     \\
Travel Cost * (Annual Pass == 0), units: 0.01 CHF (Train)      &      -1.6303  &        0.096     &   -17.055  &         0.000**        &       -1.818    &       -1.443     \\
Travel Cost * (Annual Pass == 0), units: 0.01 CHF (Swissmetro) &      -0.7797  &        0.047     &   -16.458  &         0.000**        &       -0.872    &       -0.687     \\
Travel Cost, units: 0.01 CHF (Car)                             &      -1.0766  &        0.116     &    -9.271  &         0.000**        &       -1.304    &       -0.849     \\
Headway, units:hrs, (Train)                                    &      -0.3296  &        0.065     &    -5.102  &         0.000**        &       -0.456    &       -0.203     \\
Headway, units:hrs, (Swissmetro)                               &      -0.5052  &        0.203     &    -2.484  &         0.013*        &       -0.904    &       -0.107     \\
Airline Seat Configuration, base=No (Swissmetro)               &      -0.5768  &        0.093     &    -6.192  &         0.000**        &       -0.759    &       -0.394     \\
Surveyed on a Train, base=No, (Train and Swissmetro           &       3.0548  &        0.141     &    21.732  &         0.000**        &        2.779    &        3.330     \\
First Class == False, (Swissmetro)                             &       0.0408  &        0.084     &     0.483  &         0.629        &       -0.125    &        0.206     \\
Number of Luggage Pieces == 1, (Car)                          &       0.3868  &        0.070     &     5.564  &         0.000**        &        0.251    &        0.523     \\
Number of Luggage Pieces $>$ 1, (Car)                          &       2.0613  &        0.364     &     5.659  &         0.000**        &        1.347    &        2.775     \\
\bottomrule
\end{tabular}\caption{Multinomial Logit Model Regression coefficients - original model (**= p<0.05)}\label{original_2}
\end{table}
\end{center}

%%% PUT HERE

Given that embeddings allow us to be ambitious with categorical variable dimensionality, we decided to incorporate an Origin-Destination (OD) component that is available in the dataset (26$\times$26 Swiss cantons). Other new variables to extend the above specification were (explanations extracted from \cite{swissmetro_desc}):
\begin{itemize}
\item TICKET - Travel ticket. 0: None, 1: Two way with half price card, 2: One way with half price card, 3: Two way normal price, 4: One way normal price, 5: Half day, 6: Annual season ticket, 7: Annual season ticket Junior or Senior, 8: Free travel after 7pm card, 9: Group ticket, 10: Other
\item WHO - Who pays [for travel] (0: unknown, 1: self, 2: employer, 3: half-half)
\item AGE - It captures the age class of individuals. The age-class coding scheme is of the type: $1: age<=24, 2: 24<age<=39, 3: 39<age<=54, 4: 54<age<=65, 5: 65 <age, 6: not\:\: known$
\item INCOME - Traveler’s income per year [thousand CHF] 0 or 1: under 50. 2: between 50 and 100. 3: over 100. 4: unknown
\end{itemize}

From now on, we will call the set of these variables, the \emph{encoding set}. I.e. \emph{encoding set}=\{OD, TICKET, WHO, AGE, INCOME\}. All variables in the encoding set, regardless of their encoding treatment (dummy, embeddings or PCA), will have independent parameters in utility specifications for alternatives 1 and 2 (train and swissmetro, respectively). %When we included them in car (``default") alternative, we observed that models became more unstable (e.g. multicollinearity) in some variables. 

An important decision regards, for each variable in the encoding set, the number $K$ to use, in embeddings and PCA models. For the sake of comparability, we will keep the same values of $K$ for the embeddings sizes and number of principal components in PCA. Through an incremental grid search (where we started with a single variable, and then added subsequently)\footnote{We equipped PyTre for ``automatic selection" of K through 90\% variance explained rule using PCA, i.e. the number $K$ is determined by the number of eigenvectors necessary to explain 90\% of variance in each variable in the training set. The results were in generally however worse than our search, so we keep them in this paper. They were, however, generally consistent (Embeddings model gets best performance, PCA follows soon after, then shortened dummy variables model).}, we arrived to Table \ref{listofk}.

\begin{center}
\begin{table}[!htbp]
\center
\scriptsize
\begin{tabular}{lcc}
& \textbf{K} & \textbf{Original dim} \\
OD & 3 & 96\\
TICKET & 5 & 9\\
WHO& 1 &4\\
AGE & 3 & 5\\
INCOME&3&4\\
\end{tabular} \caption{New dimensionality ($K$) of encoding set variables}\label{listofk}
\end{table}
\end{center}

 \subsection{Embeddings model}

\begin{figure}[htbp]
\begin{center}
\includegraphics[width=4.65in]{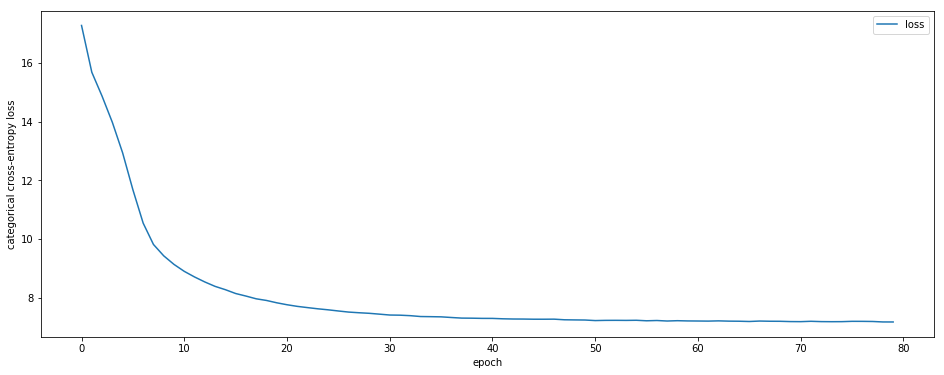}
\caption{Embeddings model training performance}\label{performance}
\end{center}
\end{figure}

It is now time to estimate and test our embeddings model as depicted in Figure \ref{TravelBehaviorEmbeddings} right, therefore all variables in the encoding set will be estimated simultaneously. As mentioned before, due to stochasticity, we repeat each experiment multiple times, and the model reported here is the one with best development set log-likelihood. We repeated the experiment 300 times. Each time, we train model through 80 epochs (requiring about 1 minute overall, in an affordable server with GPU boards). Figure \ref{performance} shows the typical performance in neural network training, where the first epochs see a dramatic improvement in objective function loss, and then the model converges approximately at a monotonic decreasing rate. It is our experience that this difference reduces substantially as dataset size grows. 

\begin{figure}[htbp!]
\begin{center}
\includegraphics[width=5in]{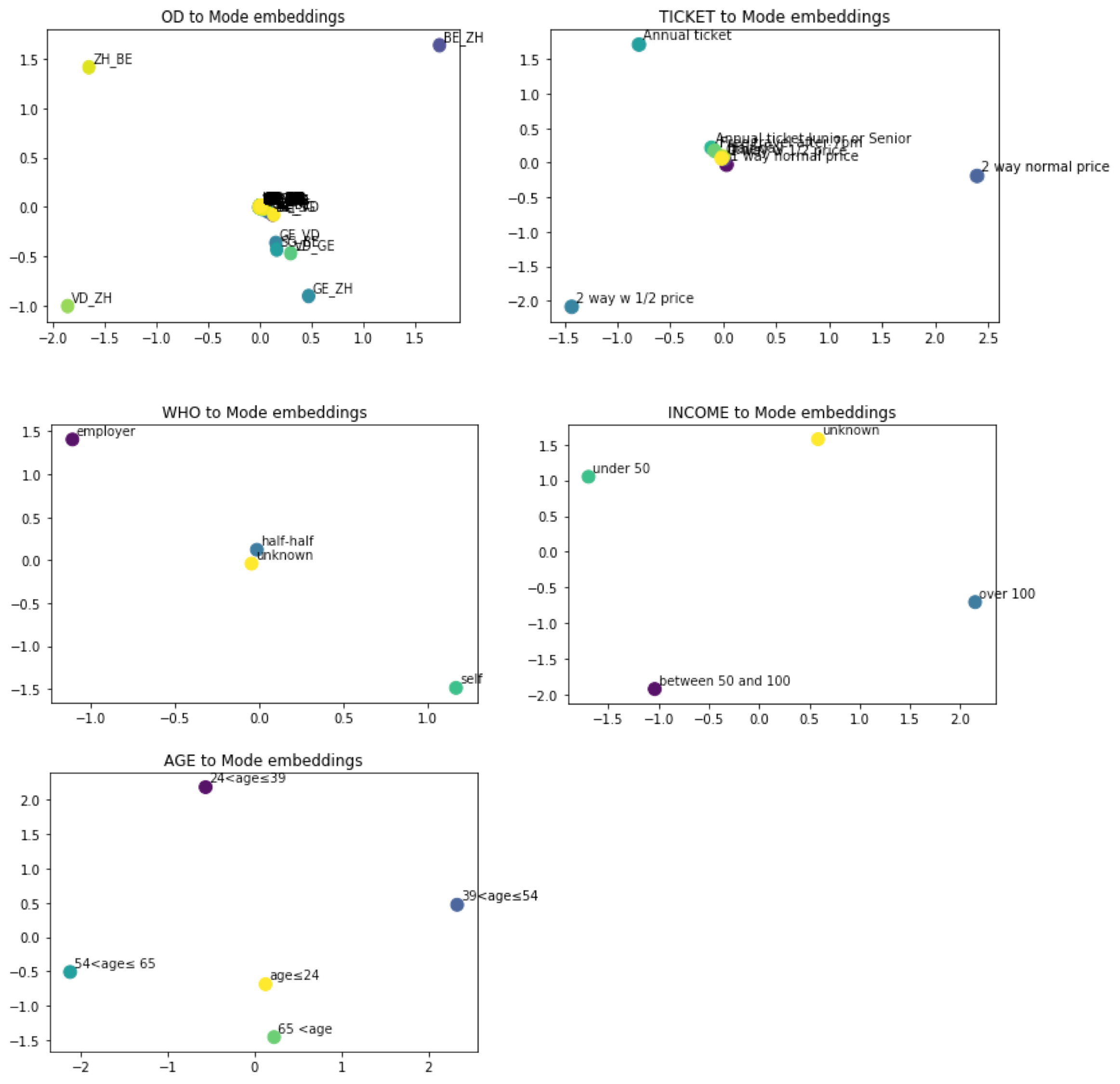}
\caption{MDS visualizations of embeddings results}\label{visualization}
\end{center}
\end{figure}

Since the dimensionality of embeddings is higher than 2, it becomes humanly impossible to directly visualize them. A common technique to use is called \emph{Multi-dimensional scaling} (MDS), which is similar to the one presented in Figure \ref{fastext_2019}. In such a visualization, data is projected in 2D space by maintaining the same vector-to-vector distances as in the original ($K$ order space). Therefore the X and Y axes have no specific meaning, only distances between every pair of points are relevant. Figure \ref{visualization} shows the MDS visualizations for all variables in the encoding set. 

\begin{figure}[htbp]
\begin{center}
\includegraphics[width=3in]{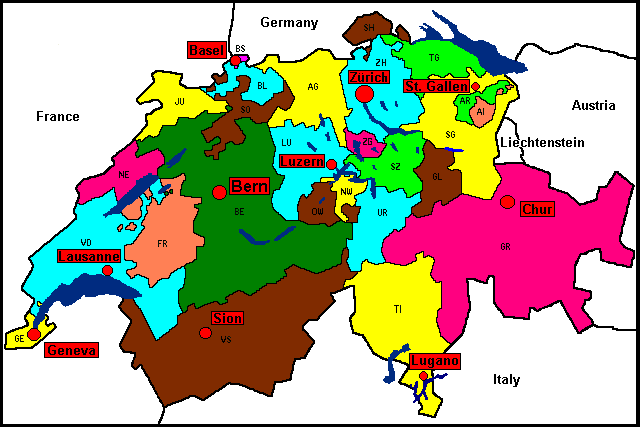}
\caption{Switzerland's cantons}\label{cantons}
\end{center}
\end{figure}

The interpretation of such visualization needs to focus on similarity between sets of categories. If two categories are very close, it means that, from the perspective of mode choice, they have a similar effect. The OD visualization shows some intuitive results: while a vast majority of OD pairs seems to have relatively similar effects, ODs between three pairs of major cities/cantons (Zurich, ZH and Berne, BE; Genève, GE, and Lausanne, VD; Lausanne, VD and Zurich, ZH) seem to have particularly different effects. To help understand the cantons and their geographical location, we show the map in Figure \ref{cantons}.

Regarding the variable TICKET, it shows that ``2-way" and normal ``annual ticket" seem to imply a different choice behavior, while to a much lesser extent ``free travel" and ``annual ticket junior or senior" seem related. Regarding who pays (WHO variable), quite intuitively ``employer" and ``self" lie in opposite ends, while ``half-half" seems falls almost right in the middle. An interesting pattern happens with the ``unknown" category, which falls in the centroid of all categories, and this happens again in AGE. A possible interpretation is that the \emph{true} value of ``unknown" are uniformly distributed across categories, which makes some sense. Regarding INCOME, all categories except for ``unknown" are approximately equidistant, which implies little advantage with respect to a dummy variable encoding.  

We now present the results of our mode choice model with embeddings. Notice that the number of parameters will be higher than the simpler model presented before. The original one has 14 variables, but it now grows up to 39. This is still quite small, if we consider the fully expanded dummy variables specification (232 variables). Table \ref{emb_results} shows the summary of results. To let the reader look into the model, we show an excerpt of the variables in Table \ref{embeddings_original}, but since this is not directly interpretable, we projected back the encodings into the original categorical space (Table \ref{embeddings_projected}), i.e. we calculated the coefficient for each potential dummy variable, together with its p-values, as mentioned in section \ref{concept}. Notice that the dimensionality is now very high (232 variables), particularly due to so many ``OD" variables. therefore we only show an excerpt of those coefficients (only those that are both statistically significant and with an absolute value greater than 0.05). We show the full set of coefficients for all other embeddings variables. This model was capable of estimating statistically significant coefficients to 164 OD pairs (out of 178), giving and overall count of 188 statistically significant coefficients (out of 232). 

\begin{center}
\begin{table}[!htbp]
\center
\scriptsize
\begin{tabular}{lclc}
\toprule
Dep. Variable:                                                 &          CHOICE         & No. Observations: &   6,373     \\
Model:                                                         & MNL & Df Residuals:    &   6,334     \\
Method:                                                        &           MLE           & Df Model:        &     39      \\
Date:                                                          &     29 Aug 2019    & Pseudo R-squ.:   &   0.330     \\
Time:                                                         &         00:46:49        & Pseudo R-bar-squ.:  &   0.324     \\
AIC:                                                           &        8,871.975        & Log-Likelihood:  & -4,396.987  \\
\textbf{Pseudo R-squ. (testset):} &   \textbf{0.326}      & \textbf{ Pseudo R-bar-squ. (testset):} &   \textbf{0.312}    \\
\textbf{AIC (testset):}                                                           &        3,076.3       & Log-Likelihood (testset):  & \textbf{-1,389.9} \\
\bottomrule
\end{tabular} \caption{Testset results for embeddings model}\label{emb_results}
\end{table}
\begin{table}[!htbp]
\scriptsize
\begin{tabular}{lccHcHH}
                                                                        & \textbf{coef} & \textbf{std err} & \textbf{z} & \textbf{P$>$$|$z$|$} & \textbf{[0.025} & \textbf{0.975]}  \\
\midrule
ASC Train                                                      &      -1.3528  &        0.188     &    -7.204  &         0.000**        &       -1.721    &       -0.985     \\
ASC Swissmetro                                                 &      -0.3656  &        0.132     &    -2.777  &         0.005**        &       -0.624    &       -0.108     \\
Travel Time, units:hrs (Train and Swissmetro)                  &      -0.6657  &        0.045     &   -14.863  &         0.000**        &       -0.753    &       -0.578     \\
Travel Time, units:hrs (Car)                                   &      -0.7785  &        0.051     &   -15.229  &         0.000**        &       -0.879    &       -0.678     \\
Travel Cost * (Annual Pass == 0), units: 0.01 CHF (Train)      &      -1.2294  &        0.103     &   -11.955  &         0.000**        &       -1.431    &       -1.028     \\
Travel Cost * (Annual Pass == 0), units: 0.01 CHF (Swissmetro) &      -0.8755  &        0.050     &   -17.411  &         0.000**        &       -0.974    &       -0.777     \\
Travel Cost, units: 0.01 CHF (Car)                             &      -1.1659  &        0.122     &    -9.527  &         0.000**        &       -1.406    &       -0.926     \\
Headway, units:hrs, (Train)                                    &      -0.4269  &        0.070     &    -6.060  &         0.000**        &       -0.565    &       -0.289     \\
Headway, units:hrs, (Swissmetro)                               &      -0.6240  &        0.207     &    -3.014  &         0.003**        &       -1.030    &       -0.218     \\
Airline Seat Configuration, base=No (Swissmetro)               &      -0.4594  &        0.101     &    -4.571  &         0.000**        &       -0.656    &       -0.262     \\
Surveyed on a Train, base=No, (Train and Swissmetro           &       2.7057  &        0.140     &    19.286  &         0.000**        &        2.431    &        2.981     \\
First Class == False, (Swissmetro)                             &       0.1743  &        0.106     &     1.644  &         0.100        &       -0.034    &        0.382     \\
Number of Luggage Pieces == 1, (Car)                          &       0.4019  &        0.071     &     5.625  &         0.000**        &        0.262    &        0.542     \\
Number of Luggage Pieces $>$ 1, (Car)                          &       1.9096  &        0.374     &     5.101  &         0.000**        &        1.176    &        2.643     \\
OD0\_Train                                                     &      -0.0874  &        0.061     &    -1.431  &         0.152        &       -0.207    &        0.032     \\
OD0\_SM                                                        &      -0.0215  &        0.042     &    -0.515  &         0.606        &       -0.103    &        0.060     \\
OD1\_Train                                                     &      -0.4837  &        0.105     &    -4.629  &         0.000**        &       -0.689    &       -0.279     \\
OD1\_SM                                                        &      -0.5679  &        0.087     &    -6.501  &         0.000**        &       -0.739    &       -0.397     \\
OD2\_Train                                                     &      -0.4843  &        0.103     &    -4.721  &         0.000**        &       -0.685    &       -0.283     \\
OD2\_SM                                                        &      -0.6334  &        0.084     &    -7.579  &         0.000**        &       -0.797    &       -0.470     \\
TICKET0\_Train                                                 &      -2.2058  &        2.547     &    -0.866  &         0.387        &       -7.199    &        2.787     \\
TICKET1\_Train                                                 &     -69.8745  &       13.866     &    -5.039  &         0.000**        &      -97.052    &      -42.697     \\
TICKET2\_Train                                                 &      79.4235  &       35.699     &     2.225  &         0.026*        &        9.456    &      149.391     \\
TICKET3\_Train                                                 &     -18.8898  &       21.044     &    -0.898  &         0.369        &      -60.136    &       22.356     \\
TICKET4\_Train                                                 &     160.5108  &       34.763     &     4.617  &         0.000**        &       92.376    &      228.646     \\
WHO0\_Train                                                    &      -0.0390  &        0.060     &    -0.650  &         0.516        &       -0.157    &        0.079     \\
WHO0\_SM                                                       &       0.2110  &        0.037     &     5.760  &         0.000**        &        0.139    &        0.283     \\
AGE0\_Train                                                    &       0.2745  &        0.081     &     3.371  &         0.001**        &        0.115    &        0.434     \\
AGE0\_SM                                                       &      -0.2245  &        0.067     &    -3.337  &         0.001**        &       -0.356    &       -0.093     \\
AGE1\_Train                                                    &      -0.1461  &        0.081     &    -1.794  &         0.073*        &       -0.306    &        0.014     \\
AGE1\_SM                                                       &       0.1760  &        0.069     &     2.551  &         0.011**        &        0.041    &        0.311     \\
AGE2\_Train                                                    &      -0.1870  &        0.054     &    -3.446  &         0.001**        &       -0.293    &       -0.081     \\
AGE2\_SM                                                       &       0.0175  &        0.036     &     0.491  &         0.623        &       -0.052    &        0.087     \\
INCOME0\_Train                                                 &      -0.3190  &        0.058     &    -5.510  &         0.000**        &       -0.432    &       -0.206     \\
INCOME0\_SM                                                    &       0.0670  &        0.047     &     1.433  &         0.152        &       -0.025    &        0.159     \\
INCOME1\_Train                                                 &      -0.1940  &        0.073     &    -2.642  &         0.008**        &       -0.338    &       -0.050     \\
INCOME1\_SM                                                    &      -0.0436  &        0.058     &    -0.749  &         0.454        &       -0.158    &        0.071     \\
INCOME2\_Train                                                 &      -0.0498  &        0.069     &    -0.727  &         0.467        &       -0.184    &        0.084     \\
INCOME2\_SM                                                    &       0.0672  &        0.056     &     1.193  &         0.233        &       -0.043    &        0.178     \\
\end{tabular} \caption{Multinomial Logit Model Regression Results - embeddings model (* = p<0.1; ** = p<0.05)}\label{embeddings_original}
\end{table}

\begin{table}[!htbp]
\begin{center}
\scriptsize
\begin{tabular}{lccHc}
                                                                        & \textbf{coef} & \textbf{std err} & \textbf{z} & \textbf{P$>$$|$z$|$} \\
\midrule
%OD\_AG\_GE\_Train&-0.004&0.001&-5.174&0.000**\\
%OD\_AG\_SG\_Train&0.004&0.001&6.269&0.000**\\
%OD\_AG\_VD\_Train&0.000&0.000&2.930&0.003**\\
%OD\_AR\_BE\_Train&-0.002&0.001&-2.997&0.003**\\
%OD\_AR\_ZH\_Train&-0.003&0.001&-4.712&0.000**\\
%OD\_BE\_AG\_Train&0.001&0.000&1.749&0.080*\\
%OD\_BE\_AR\_Train&-0.003&0.001&-2.797&0.005**\\
%OD\_BE\_GE\_Train&-0.007&0.002&-2.955&0.003**\\
%OD\_BE\_GR\_Train&0.002&0.000&5.466&0.000**\\
OD\_BE\_SG\_Train&-0.109&0.022&-5.075&0.000**\\
%OD\_BE\_SH\_Train&-0.002&0.001&-2.821&0.005**\\
%OD\_BE\_SZ\_Train&0.001&0.000&6.410&0.000**\\
%OD\_BE\_TH\_Train&0.003&0.001&3.305&0.001**\\
%OD\_BE\_TI\_Train&0.001&0.000&3.471&0.001**\\
%OD\_BE\_VD\_Train&-0.009&0.004&-2.570&0.010**\\
%OD\_BE\_VS\_Train&0.002&0.000&5.512&0.000**\\
%OD\_BS\_VD\_Train&0.002&0.000&6.177&0.000**\\
%OD\_FR\_GE\_Train&0.005&0.001&3.981&0.000**\\
%OD\_FR\_SG\_Train&0.000&0.000&5.338&0.000**\\
%OD\_FR\_TH\_Train&-0.001&0.000&-4.095&0.000**\\
%OD\_FR\_ZH\_Train&-0.000&0.000&-6.697&0.000**\\
%OD\_GE\_AG\_Train&0.006&0.001&5.228&0.000**\\
%OD\_GE\_AI\_Train&-0.004&0.001&-4.738&0.000**\\
%OD\_GE\_BL\_Train&-0.001&0.000&-4.445&0.000**\\
%OD\_GE\_BS\_Train&-0.007&0.001&-4.712&0.000**\\
%OD\_GE\_FR\_Train&-0.007&0.002&-3.224&0.001**\\
%OD\_GE\_GR\_Train&-0.006&0.001&-5.333&0.000**\\
%OD\_GE\_LU\_Train&-0.006&0.001&-5.467&0.000**\\
%OD\_GE\_SH\_Train&-0.004&0.001&-5.728&0.000**\\
%OD\_GE\_SO\_Train&0.001&0.000&4.819&0.000**\\
%OD\_GE\_TI\_Train&0.003&0.000&6.065&0.000**\\
OD\_GE\_VD\_Train&0.171&0.029&5.927&0.000**\\
%OD\_GE\_ZG\_Train&0.001&0.000&3.713&0.000**\\
OD\_GE\_ZH\_Train&-0.416&0.084&-4.961&0.000**\\
%OD\_GL\_FR\_Train&0.001&0.000&6.601&0.000**\\
%OD\_GR\_BE\_Train&-0.004&0.001&-4.026&0.000**\\
%OD\_GR\_GE\_Train&0.002&0.000&5.202&0.000**\\
%OD\_GR\_VD\_Train&-0.006&0.001&-5.369&0.000**\\
%OD\_LU\_VD\_Train&0.001&0.000&1.770&0.077*\\
%OD\_SG\_AG\_Train&0.001&0.000&5.681&0.000**\\
OD\_SG\_BE\_Train&0.065&0.024&2.670&0.008**\\
%OD\_SG\_BL\_Train&-0.002&0.001&-3.057&0.002**\\
%OD\_SG\_BS\_Train&-0.002&0.000&-4.737&0.000**\\
%OD\_SG\_FR\_Train&-0.004&0.001&-5.043&0.000**\\
%OD\_SG\_GE\_Train&0.006&0.001&4.013&0.000**\\
%OD\_SG\_SO\_Train&0.002&0.000&4.717&0.000**\\
%OD\_SG\_VD\_Train&0.004&0.001&3.318&0.001**\\
%OD\_SG\_ZH\_Train&0.012&0.004&3.326&0.001**\\
%OD\_SH\_BE\_Train&-0.001&0.000&-4.626&0.000**\\
%OD\_TH\_BE\_Train&0.003&0.001&5.295&0.000**\\
%OD\_TH\_FR\_Train&-0.002&0.000&-4.024&0.000**\\
%OD\_TH\_GE\_Train&-0.003&0.001&-5.002&0.000**\\
%OD\_TH\_VD\_Train&0.001&0.000&2.827&0.005**\\
%OD\_TH\_ZH\_Train&0.001&0.000&6.031&0.000**\\
%OD\_TI\_SG\_Train&0.002&0.000&5.535&0.000**\\
%OD\_VD\_AG\_Train&-0.002&0.001&-2.854&0.004**\\
%OD\_VD\_BE\_Train&-0.011&0.005&-2.356&0.018**\\
OD\_VD\_GE\_Train&0.660&0.099&6.675&0.000**\\
%OD\_VD\_GR\_Train&-0.002&0.001&-3.161&0.002**\\
%OD\_VD\_LU\_Train&-0.000&0.000&-1.979&0.048**\\
%OD\_VD\_SH\_Train&0.003&0.001&5.780&0.000**\\
%OD\_VD\_SO\_Train&-0.002&0.000&-4.736&0.000**\\
%OD\_VD\_SZ\_Train&0.002&0.000&5.862&0.000**\\
%OD\_VD\_TH\_Train&0.004&0.001&5.436&0.000**\\
%OD\_VD\_TI\_Train&-0.005&0.001&-5.337&0.000**\\
%OD\_VD\_ZG\_Train&0.009&0.002&5.114&0.000**\\
%OD\_VS\_GE\_Train&0.004&0.001&4.411&0.000**\\
%OD\_VS\_ZH\_Train&0.004&0.001&5.660&0.000**\\
%OD\_ZG\_FR\_Train&-0.003&0.001&-4.732&0.000**\\
%OD\_ZG\_GE\_Train&0.008&0.001&5.678&0.000**\\
%OD\_ZG\_SG\_Train&0.001&0.000&5.589&0.000**\\
%OD\_ZG\_VD\_Train&0.003&0.000&6.090&0.000**\\
%OD\_ZH\_AR\_Train&-0.002&0.001&-2.185&0.029**\\
%OD\_ZH\_FR\_Train&0.007&0.001&5.830&0.000**\\
%OD\_ZH\_SG\_Train&-0.007&0.004&-1.742&0.081*\\
%OD\_ZH\_TH\_Train&0.003&0.001&5.573&0.000**\\
%OD\_ZH\_VD\_Train&-0.048&0.014&-3.402&0.001**\\
%OD\_ZH\_VS\_Train&-0.011&0.002&-5.103&0.000**\\
%OD\_AG\_GE\_SM&-0.006&0.001&-8.944&0.000**\\
%OD\_AG\_SG\_SM&0.006&0.001&9.623&0.000**\\
%OD\_AG\_VD\_SM&0.001&0.000&5.297&0.000**\\
%OD\_AG\_ZH\_SM&-0.001&0.001&-1.867&0.062*\\
%OD\_AR\_BE\_SM&-0.003&0.001&-5.836&0.000**\\
%OD\_AR\_ZH\_SM&-0.005&0.001&-8.244&0.000**\\
%OD\_BE\_AR\_SM&-0.005&0.001&-4.947&0.000**\\
%OD\_BE\_GE\_SM&-0.010&0.002&-5.478&0.000**\\
%OD\_BE\_GR\_SM&0.002&0.000&7.547&0.000**\\
OD\_BE\_SG\_SM&-0.149&0.017&-8.597&0.000**\\
%OD\_BE\_SH\_SM&-0.003&0.000&-5.809&0.000**\\
%OD\_BE\_SZ\_SM&0.001&0.000&9.829&0.000**\\
%OD\_BE\_TH\_SM&0.004&0.001&4.567&0.000**\\
%OD\_BE\_TI\_SM&0.001&0.000&4.536&0.000**\\
%OD\_BE\_VD\_SM&-0.015&0.003&-5.280&0.000**\\
%OD\_BE\_VS\_SM&0.002&0.000&8.176&0.000**\\
%OD\_BS\_VD\_SM&0.002&0.000&9.242&0.000**\\
%OD\_FR\_GE\_SM&0.006&0.001&5.553&0.000**\\
%OD\_FR\_SG\_SM&0.000&0.000&9.049&0.000**\\
%OD\_FR\_TH\_SM&-0.002&0.000&-7.379&0.000**\\
%OD\_FR\_ZH\_SM&-0.001&0.000&-9.633&0.000**\\
%OD\_GE\_AG\_SM&0.008&0.001&8.966&0.000**\\
%OD\_GE\_AI\_SM&-0.005&0.001&-8.551&0.000**\\
%OD\_GE\_BE\_SM&-0.009&0.006&-1.680&0.093*\\
%OD\_GE\_BL\_SM&-0.001&0.000&-8.033&0.000**\\
%OD\_GE\_BS\_SM&-0.010&0.001&-8.373&0.000**\\
%OD\_GE\_FR\_SM&-0.011&0.002&-5.955&0.000**\\
%OD\_GE\_GR\_SM&-0.008&0.001&-9.081&0.000**\\
%OD\_GE\_LU\_SM&-0.008&0.001&-9.194&0.000**\\
%OD\_GE\_SG\_SM&-0.001&0.001&-2.104&0.035**\\
%OD\_GE\_SH\_SM&-0.005&0.001&-9.346&0.000**\\
%OD\_GE\_SO\_SM&0.001&0.000&8.588&0.000**\\
%OD\_GE\_TH\_SM&-0.000&0.000&-1.967&0.049**\\
%OD\_GE\_TI\_SM&0.004&0.000&9.452&0.000**\\
OD\_GE\_VD\_SM&0.196&0.024&8.239&0.000**\\
%OD\_GE\_VS\_SM&-0.003&0.001&-3.155&0.002**\\
%OD\_GE\_ZG\_SM&0.001&0.000&6.074&0.000**\\
OD\_GE\_ZH\_SM&-0.564&0.068&-8.306&0.000**\\
%OD\_GL\_FR\_SM&0.001&0.000&9.920&0.000**\\
%OD\_GR\_BE\_SM&-0.005&0.001&-7.269&0.000**\\
%OD\_GR\_GE\_SM&0.003&0.000&8.458&0.000**\\
%OD\_GR\_VD\_SM&-0.009&0.001&-9.114&0.000**\\
%OD\_LU\_VD\_SM&0.000&0.000&2.073&0.038**\\
%OD\_SG\_AG\_SM&0.001&0.000&9.040&0.000**\\
OD\_SG\_BE\_SM&0.060&0.020&3.038&0.002**\\
%OD\_SG\_BL\_SM&-0.002&0.000&-5.853&0.000**\\
%OD\_SG\_BS\_SM&-0.003&0.000&-8.048&0.000**\\
%OD\_SG\_FR\_SM&-0.006&0.001&-8.741&0.000**\\
%OD\_SG\_GE\_SM&0.007&0.001&5.680&0.000**\\
%OD\_SG\_SO\_SM&0.003&0.000&8.518&0.000**\\
%OD\_SG\_TH\_SM&-0.001&0.000&-2.718&0.007**\\
%OD\_SG\_VD\_SM&0.004&0.001&4.415&0.000**\\
%OD\_SG\_ZH\_SM&0.013&0.003&4.421&0.000**\\
%OD\_SH\_BE\_SM&-0.001&0.000&-7.617&0.000**\\
%OD\_TH\_BE\_SM&0.004&0.000&8.314&0.000**\\
%OD\_TH\_FR\_SM&-0.003&0.000&-7.612&0.000**\\
%OD\_TH\_GE\_SM&-0.004&0.000&-8.615&0.000**\\
%OD\_TH\_VD\_SM&0.001&0.000&3.488&0.000**\\
%OD\_TH\_ZH\_SM&0.001&0.000&9.535&0.000**\\
%OD\_TI\_SG\_SM&0.002&0.000&8.932&0.000**\\
%OD\_VD\_AG\_SM&-0.004&0.001&-5.922&0.000**\\
%OD\_VD\_BE\_SM&-0.017&0.004&-4.529&0.000**\\
OD\_VD\_GE\_SM&0.803&0.082&9.817&0.000**\\
%OD\_VD\_GR\_SM&-0.002&0.000&-5.748&0.000**\\
%OD\_VD\_LU\_SM&-0.001&0.000&-3.905&0.000**\\
%OD\_VD\_SG\_SM&-0.005&0.002&-2.512&0.012**\\
%OD\_VD\_SH\_SM&0.004&0.000&9.279&0.000**\\
%OD\_VD\_SO\_SM&-0.003&0.000&-8.532&0.000**\\
%OD\_VD\_SZ\_SM&0.002&0.000&9.006&0.000**\\
%OD\_VD\_TH\_SM&0.004&0.001&8.086&0.000**\\
%OD\_VD\_TI\_SM&-0.006&0.001&-9.053&0.000**\\
%OD\_VD\_ZG\_SM&0.011&0.001&8.165&0.000**\\
%OD\_VS\_BE\_SM&-0.000&0.000&-2.602&0.009**\\
%OD\_VS\_GE\_SM&0.005&0.001&6.725&0.000**\\
%OD\_VS\_ZH\_SM&0.005&0.001&8.892&0.000**\\
%OD\_ZG\_FR\_SM&-0.004&0.000&-8.547&0.000**\\
%OD\_ZG\_GE\_SM&0.010&0.001&9.048&0.000**\\
%OD\_ZG\_SG\_SM&0.002&0.000&9.185&0.000**\\
%OD\_ZG\_VD\_SM&0.003&0.000&9.648&0.000**\\
%OD\_ZH\_AR\_SM&-0.003&0.001&-4.184&0.000**\\
%OD\_ZH\_FR\_SM&0.009&0.001&8.967&0.000**\\
%OD\_ZH\_GE\_SM&-0.007&0.002&-2.910&0.004**\\
%OD\_ZH\_SG\_SM&-0.012&0.003&-3.575&0.000**\\
%OD\_ZH\_TH\_SM&0.004&0.000&8.478&0.000**\\
OD\_ZH\_VD\_SM&-0.070&0.011&-6.184&0.000**\\
%OD\_ZH\_VS\_SM&-0.015&0.002&-8.733&0.000**\\
WHO\_half-half\_Train&-0.011&0.017&-0.650&0.516\\
WHO\_self\_Train&0.070&0.107&0.650&0.516\\
WHO\_unknown\_Train&-0.005&0.008&-0.650&0.516\\
WHO\_half-half\_SM&0.059&0.010&5.760&0.000**\\
WHO\_self\_SM&-0.376&0.065&-5.760&0.000**\\
WHO\_unknown\_SM&0.028&0.005&5.760&0.000**\\
TICKET\_1 way w 1/2 price\_Train&-0.008&1.676&-0.005&0.996\\
TICKET\_2 way normal price\_Train&12.259&61.787&0.198&0.843\\
TICKET\_2 way w 1/2 price\_Train&-10.791&61.797&-0.175&0.861\\
TICKET\_Annual ticket\_Train&-4.824&26.095&-0.185&0.853\\
TICKET\_Annual ticket Junior or Senior\_Train&1.607&2.128&0.755&0.450\\
TICKET\_Free travel after 7pm\_Train&1.537&2.583&0.595&0.552\\
TICKET\_Half day\_Train&1.344&2.034&0.661&0.509\\
TICKET\_Other\_Train&1.050&1.200&0.875&0.382\\
INCOME\_over 100\_Train&-0.168&0.147&-1.142&0.253\\
INCOME\_under 50\_Train&0.085&0.137&0.622&0.534\\
INCOME\_unknown\_Train&0.633&0.113&5.609&0.000**\\
INCOME\_over 100\_SM&0.214&0.118&1.817&0.069*\\
INCOME\_under 50\_SM&-0.029&0.111&-0.264&0.792\\
INCOME\_unknown\_SM&-0.109&0.091&-1.197&0.231\\
AGE\_39<age<=54\_Train&-0.083&0.170&-0.490&0.624\\
AGE\_54<age<=65\_Train&0.341&0.104&3.277&0.001**\\
AGE\_65 <age\_Train&0.541&0.131&4.136&0.000**\\
AGE\_age<=24\_Train&-0.299&0.108&-2.766&0.006**\\
AGE\_39<age<=54\_SM&-0.021&0.138&-0.152&0.879\\
AGE\_54<age<=65\_SM&-0.051&0.070&-0.731&0.465\\
AGE\_65 <age\_SM&-0.460&0.109&-4.222&0.000**\\
AGE\_age<=24\_SM&0.356&0.090&3.953&0.000**\\
%$\cdots$&$\cdots$&$\cdots$&$\cdots$&$\cdots$&$\cdots$&$\cdots$\\
\bottomrule
\end{tabular} \caption{Multinomial Logit Model Regression Results - embeddings model projected into dummy variable space (* = p<0.1; ** = p<0.05)}\label{embeddings_projected}
\end{center}
\end{table}

\end{center}

Unsurprisingly, this model outperforms considerably the original one, as it uses more data. A fairer comparison will be made with the dummy and PCA baseline models. Regarding the estimated coefficients, all signs remained consistent with the original model, while the magnitudes varied slightly, which is not surprising. 

We can also verify that general signs for the WHO variable (\emph{who pays?}) make sense, particularly considering that the base category is ``employer". Unless the employer pays at least half, people prefer to take the (cheaper) train. This is also consistent with the INCOME variable, where lower income people clearly prefer train to swissmetro and higher income people prefer swissmetro, even though the coefficients are not all statistically significant. Regarding the AGE variable (base category is 24 to 39 years old), it seems younger people prefer SM, while older people prefer train. On the other hand, the TICKET variable shows somewhat problematic results, with very high standard errors, and no significant coefficient. 

It is quite notorious that with only a dimensionality of 3, for the OD categorical variable (from an original number of 178 dummy variables), the model is capable of estimating reliable coefficients to almost all of the categories, while the corresponding dummy variable version fails completely. This is coherent with the belief that using shared latent spaces brings gains in variable estimation. This matches our intuition that, by sharing the latent space, categorical variables virtually use all dataset in the estimation, differently to the dummy encoding. 

%Regarding the OD variables, it is notorious that the number of statistically significant coefficients is quite small (13) comparing with the large number of such variables, and none of the ``big cities" ODs mentioned before lead to a significant coefficient. As we will see below, the dummy variables model will get zero significant OD variables, while PCA will get 28. 

%On the other hand, removing these variables would result in general degradation of the model, both in train as well as test set performance. This ambivalent result for the OD variables led us to include a model without such variables in the overall comparison in the end of this section. 

While the embeddings model shows considerable improvement with respect to the dummy variables one, it shows much less dramatic improvements in comparison with PCA. This general pattern was seen across all our experiments with this dataset, including when we vary the value of $K$/embeddings size. We will return to this subject later on.

%Finally, before moving on to the baseline models, we need to emphasize that, in an embeddings context, dataset size is very important. 

\subsection{Baseline model: dummy variable encoding}
%The first baseline model uses dummy variables for the encoding set, so each categorical variable is replaced by N-1 dummy variables. %After standardizing or normalizing the data, the (training and test set likelihood) results vary minimally, so we will keep the original values for simplicity in this paper. 

The first baseline model uses dummy variables for the encoding set, so all categorical variables, including the encoding set were assigned $D-1$ dummy variables\footnote{One could argue also for comparing embeddings with contrasting methods instead. The experiments showed negligible difference of dummies versus the other contrasting methods, therefore, for consistency with literature, we continued with dummy variables.}, where $D$ is the number of different categories in the variable. This yields a total of 232 variables. Table \ref{dummy_fit} shows a summary of the model results. It is quite clear and unsurprising that this model has estimation problems, given its complexity and the relatively small size of the dataset. The standard errors of the constants and several variables show difficulty in converging to stable values, in fact no OD dummy variable obtained a statistical significant coefficient. As is well-known, further explorations in reducing the specification could lead to better results eventually in terms of coefficients, of course degrading the estimation fit. As mentioned before, this is an exercise we will avoid in this paper. Regarding the test set results, the (pseudo) R-square beats the original model, but it is at the expense of an over-specification, as can be seen in the robust R-square result. 

We thus decided to specify a more robust dummy variables model, that excludes the OD variables. Regardless, for the sake of completeness, we will include both models later in the comparison table. As can be seen in Table \ref{dummy_reduced_fit}, the results for a model without OD variables are substantially more reasonable, and we show the coefficients in Table \ref{dummy_reduced_coeff}. The variable for ``First Class == False, (Swissmetro)" is inconsistent with the original model, but it is statistically non-significant. The alternative specific constants now have much higher standard errors and high p values, but all the others variables from the original model seem to be consistent in terms of signal and magnitude. 

Regarding the encoding set variables (excluding OD), the results do not fall behind the embeddings model in terms of statistical significant. In fact, we now have 13 significant coefficients, as opposed to 10 in the embeddings model. Curiously, it is the TICKET variable that makes the difference, perhaps indicating that a combined model (dummies for TICKET, embeddings for the others) could be a good solution.  

%Unsurprisingly, a less complex model becomes more stable, even though far from perfect (e.g. one non-significant ASC). 

%One could certainly go other ways to improve the model, for example using interactions, 

%, while table \ref{dummy_coeff} presents a small excerpt of the resulting coefficients

%Certainly, there should exist intermediate models that would also be acceptable, but they are only possible by cherrypicking individual variables, which is in itself a questionable and somewhat arbitrary exercise, certainly for this paper. 

\begin{center}

\begin{table}[!htbp]
\center
\scriptsize
\begin{tabular}{lclc}
\toprule
Dep. Variable:                                                 &          CHOICE         & No. Observations: &   6,373     \\
Model:                                                         & MNL & Df Residuals:    &   6,141     \\
Method:                                                        &           MLE           & Df Model:        &    232      \\
Date:                                                          &     Thu, 29 Aug 2019    & Pseudo R-squ.:   &   0.397     \\
Time:                                                         &         21:21:01        & Pseudo R-bar-squ.: &   0.361     \\
AIC:                                                           &        8,377.890        & Log-Likelihood:  & -3,956.945  \\
\textbf{Pseudo R-squ. (testset):} &   \textbf{-0.789}      & \textbf{ Pseudo R-bar-squ. (testset):} &  -1.02     \\
\textbf{AIC (testset):}                                                           &        7,846.2        & Log-Likelihood (testset):  & \textbf{-3,691.0} \\
%Dep. Variable:                                                 &          MODE         & No. Observations: &   7,837     \\
%Model:                                                         & MNL & Df Residuals:    &   7,787     \\
%Method:                                                        &           MLE           & Df Model:        &     50      \\
%Date:                                                          &     Thu, 27 Jun 2019    & Pseudo R-squ.:   &   0.361     \\
%Time:                                                         &         23:04:01        & Pseudo R-bar-squ.: } &   0.355     \\
%AIC:                                                           &        10.438.871       & Log-Likelihood:  & -5,169.436  \\
%\textbf{Pseudo R-squ. (testset):} &   \textbf{0.288}      & \textbf{ Pseudo R-bar-squ. (testset):} &   \textbf{0.273}     \\
%\textbf{AIC (testset):}                                                           &        \textbf{3,736.6}        & Log-Likelihood (testset):  & \textbf{-1818.3} \\

\bottomrule
\end{tabular}
\caption{Multinomial Logit Model Regression Results for dummy variable model \textbf{with} OD variables}\label{dummy_fit}
\end{table}

\begin{table}[!htbp]
\center
\scriptsize
\begin{tabular}{lclc}
\toprule
%Dep. Variable:                                                 &          MODE         & No. Observations: &   7,837     \\
%Model:                                                         & MNL & Df Residuals:    &   7,607     \\
%Method:                                                        &           MLE           & Df Model:        &    230      \\
%Date:                                                          &     Wed, 26 Jun 2019    & Pseudo R-squ.:   &   0.393     \\
%Time:                                                         &         23:00:45        & Pseudo R-bar-squ.: } &   0.365     \\
%AIC:                                                           &        10.284.489       & Log-Likelihood:  & -4,912.244  \\
%\textbf{Pseudo R-squ. (testset):} &   \textbf{0.070}      & \textbf{ Pseudo R-bar-squ. (testset):} &   \textbf{0.147}     \\
%\textbf{AIC (testset):}                                                           &        \textbf{5,207.4}        & Log-Likelihood (testset):  & \textbf{-2373.7} \\
Dep. Variable:                                                 &          CHOICE         & No. Observations: &   6,373     \\
Model:                                                         & MNL & Df Residuals:    &   6,331     \\
Method:                                                        &           MLE           & Df Model:        &     42      \\
Date:                                                          &     Thu, 29 Aug 2019    & Pseudo R-squ.:   &   0.331     \\
Time:                                                         &         21:25:10        & Pseudo R-bar-squ.:  &   0.325     \\
AIC:                                                           &        8,857.848        & Log-Likelihood:  & -4,386.924  \\
\textbf{Pseudo R-squ. (testset):} &   \textbf{0.307}      & \textbf{ Pseudo R-bar-squ. (testset):} &  0.292     \\
\textbf{AIC (testset):}                                                           &       2,942.5       & Log-Likelihood (testset):  & \textbf{-1,429.3} \\

\bottomrule
\end{tabular}
\caption{Multinomial Logit Model Regression Results for dummy variable model \textbf{without} OD variables}\label{dummy_reduced_fit}
\end{table}

\begin{table}[!htbp]
\scriptsize
\begin{tabular}{lccHcHH}
                                                                        & \textbf{coef} & \textbf{std err} & \textbf{z} & \textbf{P$>$$|$z$|$} & \textbf{[0.025} & \textbf{0.975]}  \\
\midrule
ASC Train                                                      &      -2.1976  &        0.464     &    -4.732  &         0.000**        &       -3.108    &       -1.287     \\
ASC Swissmetro                                                 &      -0.0183  &        0.147     &    -0.125  &         0.901        &       -0.306    &        0.269     \\
Travel Time, units:hrs (Train and Swissmetro)                  &      -0.5777  &        0.043     &   -13.481  &         0.000**        &       -0.662    &       -0.494     \\
Travel Time, units:hrs (Car)                                   &      -0.7028  &        0.049     &   -14.399  &         0.000**        &       -0.798    &       -0.607     \\
Travel Cost * (Annual Pass == 0), units: 0.01 CHF (Train)      &      -1.3984  &        0.144     &    -9.717  &         0.000**        &       -1.680    &       -1.116     \\
Travel Cost * (Annual Pass == 0), units: 0.01 CHF (Swissmetro) &      -0.8554  &        0.049     &   -17.328  &         0.000**        &       -0.952    &       -0.759     \\
Travel Cost, units: 0.01 CHF (Car)                             &      -0.9836  &        0.118     &    -8.368  &         0.000**        &       -1.214    &       -0.753     \\
Headway, units:hrs, (Train)                                    &      -0.4318  &        0.071     &    -6.042  &         0.000**        &       -0.572    &       -0.292     \\
Headway, units:hrs, (Swissmetro)                               &      -0.5994  &        0.207     &    -2.890  &         0.004**        &       -1.006    &       -0.193     \\
Airline Seat Configuration, base=No (Swissmetro)               &      -0.4091  &        0.102     &    -4.012  &         0.000**        &       -0.609    &       -0.209     \\
Surveyed on a Train, base=No, (Train and Swissmetro)           &       2.9172  &        0.143     &    20.364  &         0.000**        &        2.636    &        3.198     \\
First Class == False, (Swissmetro)                             &       0.0030  &        0.113     &     0.026  &         0.979        &       -0.218    &        0.224     \\
Number of Luggage Pieces == 1, (Car)                          &       0.3632  &        0.072     &     5.067  &         0.000**        &        0.223    &        0.504     \\
Number of Luggage Pieces $>$ 1, (Car)                          &       1.8804  &        0.375     &     5.017  &         0.000**        &        1.146    &        2.615     \\
TICKET\_1 way w 1/2 price\_Train                               &       0.1981  &        0.464     &     0.427  &         0.669        &       -0.711    &        1.107     \\
TICKET\_2 way normal price\_Train                              &       0.1965  &        0.434     &     0.452  &         0.651        &       -0.655    &        1.048     \\
TICKET\_2 way w 1/2 price\_Train                               &       0.6297  &        0.414     &     1.520  &         0.129        &       -0.182    &        1.442     \\
TICKET\_Annual ticket\_Train                                   &       1.5798  &        0.421     &     3.750  &         0.000**        &        0.754    &        2.405     \\
TICKET\_Annual ticket Junior or Senior\_Train                  &       1.5625  &        0.444     &     3.518  &         0.000**        &        0.692    &        2.433     \\
TICKET\_Free travel after 7pm\_Train                           &       2.2375  &        0.499     &     4.483  &         0.000**        &        1.259    &        3.216     \\
TICKET\_Half day\_Train                                        &       0.9871  &        0.480     &     2.056  &         0.040**        &        0.046    &        1.928     \\
TICKET\_Other\_Train                                           &       1.4718  &        0.470     &     3.129  &         0.002**        &        0.550    &        2.394     \\
WHO\_half-half\_Train                                          &       0.0966  &        0.196     &     0.492  &         0.622        &       -0.288    &        0.481     \\
WHO\_half-half\_SM                                             &       0.0794  &        0.120     &     0.663  &         0.507        &       -0.155    &        0.314     \\
WHO\_self\_Train                                               &      -0.0088  &        0.134     &    -0.066  &         0.947        &       -0.271    &        0.253     \\
WHO\_self\_SM                                                  &      -0.4312  &        0.079     &    -5.431  &         0.000**        &       -0.587    &       -0.276     \\
WHO\_unknown\_Train                                            &      -0.6166  &        0.508     &    -1.215  &         0.225        &       -1.612    &        0.378     \\
WHO\_unknown\_SM                                               &      -0.2768  &        0.216     &    -1.284  &         0.199        &       -0.699    &        0.146     \\
AGE\_39<age<=54\_Train                                          &       0.0892  &        0.140     &     0.635  &         0.525        &       -0.186    &        0.364     \\
AGE\_39<age<=54\_SM                                             &      -0.1258  &        0.083     &    -1.516  &         0.130        &       -0.289    &        0.037     \\
AGE\_54<age<= 65\_Train                                         &       0.2716  &        0.164     &     1.658  &         0.097*        &       -0.050    &        0.593     \\
AGE\_54<age<= 65\_SM                                            &      -0.1105  &        0.097     &    -1.139  &         0.255        &       -0.301    &        0.080     \\
AGE\_65 <age\_Train                                            &       1.1688  &        0.184     &     6.335  &         0.000**        &        0.807    &        1.530     \\
AGE\_65 <age\_SM                                               &      -0.7832  &        0.149     &    -5.260  &         0.000**        &       -1.075    &       -0.491     \\
AGE\_age<=24\_Train                                             &       0.4822  &        0.287     &     1.683  &         0.092*        &       -0.079    &        1.044     \\
AGE\_age<=24\_SM                                                &      -0.2051  &        0.260     &    -0.789  &         0.430        &       -0.715    &        0.305     \\
INCOME\_over 100\_Train                                        &       0.2151  &        0.139     &     1.543  &         0.123        &       -0.058    &        0.488     \\
INCOME\_over 100\_SM                                           &       0.2213  &        0.080     &     2.760  &         0.006**        &        0.064    &        0.379     \\
INCOME\_under 50\_Train                                        &       0.3692  &        0.171     &     2.160  &         0.031*        &        0.034    &        0.704     \\
INCOME\_under 50\_SM                                           &       0.1191  &        0.132     &     0.902  &         0.367        &       -0.140    &        0.378     \\
INCOME\_unknown\_Train                                         &       0.7454  &        0.192     &     3.885  &         0.000**        &        0.369    &        1.121     \\
INCOME\_unknown\_SM                                            &      -0.0139  &        0.152     &    -0.091  &         0.927        &       -0.312    &        0.284     \\
\bottomrule
\end{tabular}
\caption{Multinomial Logit Model Regression coefficients for dummy variable model without OD variables}\label{dummy_reduced_coeff}
\end{table}
\end{center}

%SUPER interesting effect with AGE (24 to 39)!

%but the essential question is what would the effect be of having embeddings in the model, instead of dummy variables

%EXPLAIN RELATIONSHIP OF DISTANCES (difference to Dummies and contrasting)!!

\subsection{Principal Components Analysis model}

For the next baseline, we applied PCA to each individual variable separately in the encoding set. Now, we repeat the very same idea of the embeddings, by extracting one set of eigenvectors from each categorical variable in the encoding set. Besides being an entirely different algorithm to embeddings, PCA is a non-supervised method. In other words, regardless of whether the target variable is mode choice or anything else (e.g. departure time, destination), the PCA encoding will be the same. 

We used the same values $K$ as in the embeddings, so after re-representing the categorical variables with the corresponding PCA vectors, we obtain again a model specification with 39 parameters.  Table \ref{pca_results} summarizes the results while Table \ref{pca_coeffs} presents the coefficients. For the OD variables, we only present the statistically significant ones. 

Looking at the test set results, the PCA model beats the reduced (no ODs) dummy model by a marginal amount while it is beaten by a larger margin by the embeddings model, although it is generally the best behaved model in terms of statistically significant parameters of non-OD variables (14 versus 10 in embeddings, 13 in dummies). The performance with OD variable coefficients is less impressive than the embeddings model (17 coeff. vs 164,  for the embeddings).

To let us end this section with a clearer comparative analysis, Table \ref{res_summary} summarizes the results. Again we only show significant OD coefficients with value higher than 0.05.

%An important decision regards, for each variable in the encoding set, the number $K$ of eigenvectors to use. We decided to apply the 90\% variance explained rule, i.e. the number $K$ is determined by the number of eigenvectors necessary to explain 90\% of variance in each variable. Table \ref{listofk} shows the values of $K$ for each variable in the encoding set. For the sake of consistency, we will keep the same values of $K$ for the embeddings sizes in the next section.
%
%\begin{center}
%\begin{table}[!htbp]
%\center
%\scriptsize
%\begin{tabular}{lcc}
%& \textbf{K} & \textbf{Original dim} \\
%OD & 55 & 79\\
%TICKET & 5 & 9\\
%WHO& 2 &4\\
%AGE & 3 & 6\\
%INCOME&2&5\\
%\end{tabular} \caption{New dimensionality ($K$) of encoding set variables, according to PCA, with 90\% explained variance}\label{listofk}
%\end{table}
%\end{center}
\begin{center}
\begin{table}[htbp!]
\scriptsize
\begin{tabular}{lclc}
\toprule
Dep. Variable:                                                 &          CHOICE         & No. Observations: &   6,373     \\
Model:                                                         & MNL & Df Residuals:    &   6,334     \\
Method:                                                        &           MLE           & Df Model:        &     39      \\
Date:                                                          &     Thu, 29 Aug 2019    & Pseudo R-squ.:   &   0.328     \\
Time:                                                         &         21:36:50        & Pseudo R-bar-squ.:  &   0.322     \\
AIC:                                                           &        8,893.838        & Log-Likelihood:  & -4,407.919  \\
\textbf{Pseudo R-squ. (testset):} &   \textbf{0.308}      & \textbf{ Pseudo R-bar-squ. (testset):} &  0.294     \\
\textbf{AIC (testset):}                                                           &        2,931.9       & Log-Likelihood (testset):  & \textbf{-1,427.0} \\

\bottomrule
\end{tabular} \caption{Results for PCA model}\label{pca_results}
\end{table}

\begin{table}[!htbp]
\scriptsize

\begin{tabular}{lccHcHH}
                                                                        & \textbf{coef} & \textbf{std err} & \textbf{z} & \textbf{P$>$$|$z$|$} & \textbf{[0.025} & \textbf{0.975]}  \\
\midrule
ASC Train                                                      &      -1.3234  &        0.197     &    -6.718  &         0.000**        &       -1.710    &       -0.937     \\
ASC Swissmetro                                                 &      -0.2962  &        0.129     &    -2.288  &         0.022**       &       -0.550    &       -0.042     \\
Travel Time, units:hrs (Train and Swissmetro)                  &      -0.5758  &        0.043     &   -13.338  &         0.000**        &       -0.660    &       -0.491     \\
Travel Time, units:hrs (Car)                                   &      -0.7041  &        0.049     &   -14.290  &         0.000**        &       -0.801    &       -0.608     \\
Travel Cost * (Annual Pass == 0), units: 0.01 CHF (Train)      &      -1.1695  &        0.133     &    -8.818  &         0.000**        &       -1.429    &       -0.910     \\
Travel Cost * (Annual Pass == 0), units: 0.01 CHF (Swissmetro) &      -0.8582  &        0.049     &   -17.348  &         0.000**        &       -0.955    &       -0.761     \\
Travel Cost, units: 0.01 CHF (Car)                             &      -0.9896  &        0.119     &    -8.350  &         0.000**        &       -1.222    &       -0.757     \\
Headway, units:hrs, (Train)                                    &      -0.4058  &        0.071     &    -5.749  &         0.000**        &       -0.544    &       -0.267     \\
Headway, units:hrs, (Swissmetro)                               &      -0.5876  &        0.207     &    -2.839  &         0.005**        &       -0.993    &       -0.182     \\
Airline Seat Configuration, base=No (Swissmetro)               &      -0.4192  &        0.102     &    -4.120  &         0.000**        &       -0.619    &       -0.220     \\
Surveyed on a Train, base=No, (Train and Swissmetro           &       2.9434  &        0.144     &    20.471  &         0.000**        &        2.662    &        3.225     \\
First Class == False, (Swissmetro)                             &      -0.0662  &        0.108     &    -0.615  &         0.538        &       -0.277    &        0.145     \\
Number of Luggage Pieces == 1, (Car)                          &       0.3526  &        0.071     &     4.981  &         0.000**        &        0.214    &        0.491     \\
Number of Luggage Pieces $>$ 1, (Car)                          &       1.8883  &        0.376     &     5.024  &         0.000**        &        1.152    &        2.625     \\
%OD\_AR\_BE\_Train&0.001&0.001&2.209&0.027**&&\\
%OD\_AR\_ZH\_Train&0.000&0.001&1.770&0.077*&&\\
%OD\_BE\_AG\_Train&0.001&0.001&2.163&0.031**&&\\
%OD\_BE\_GE\_Train&0.000&0.001&2.058&0.040**&&\\
%OD\_BE\_GR\_Train&0.003&0.002&1.915&0.055*&&\\
%OD\_BE\_TI\_Train&0.003&0.002&1.884&0.060*&&\\
%OD\_BE\_VD\_Train&0.001&0.001&2.152&0.031**&&\\
%OD\_BE\_ZH\_Train&0.003&0.002&1.894&0.058*&&\\
%OD\_FR\_GE\_Train&0.006&0.004&1.651&0.099*&&\\
%OD\_VD\_BS\_Train&-0.023&0.013&-1.713&0.087*&&\\
%OD\_VD\_LU\_Train&-0.027&0.015&-1.771&0.077*&&\\
%OD\_VD\_SZ\_Train&-0.030&0.016&-1.786&0.074*&&\\
OD\_ZH\_VS\_Train&0.066&0.036&1.844&0.065*&&\\
%OD\_SG\_VD\_SM&0.023&0.010&2.384&0.017**&&\\
OD\_SG\_ZH\_SM&0.333&0.126&2.650&0.008**&&\\
OD\_SH\_BE\_SM&0.148&0.062&2.399&0.016**&&\\
%OD\_VD\_BS\_SM&-0.015&0.008&-1.797&0.072*&&\\
TICKET\_1 way w 1/2 price\_Train&-0.926&0.328&-2.738&0.006**&&\\
TICKET\_2 way normal price\_Train&0.318&0.300&1.154&0.249&&\\
TICKET\_2 way w 1/2 price\_Train&-0.106&0.237&-0.332&0.740&&\\
TICKET\_Annual ticket\_Train&0.096&0.236&0.520&0.603&&\\
TICKET\_Annual ticket Junior or Senior\_Train&1.006&0.260&3.971&0.000**&&\\
TICKET\_Free travel after 7pm\_Train&-0.409&0.241&-1.582&0.114&&\\
TICKET\_Half day\_Train&-0.473&0.259&-1.718&0.086*&&\\
TICKET\_Other\_Train&-0.309&0.277&-1.017&0.309&&\\
WHO\_half-half\_Train&0.041&0.077&1.300&0.193&&\\
WHO\_self\_Train&0.021&0.067&1.198&0.231&&\\
WHO\_unknown\_Train&-0.043&0.078&0.196&0.844&&\\
WHO\_half-half\_SM&-0.146&0.047&-7.502&0.000**&&\\
WHO\_self\_SM&-0.076&0.041&-6.909&0.000**&&\\
WHO\_unknown\_SM&0.152&0.048&-1.132&0.258&&\\
AGE\_39<age<=54\_Train&0.669&0.176&3.998&0.000**&&\\
AGE\_54<age<= 65\_Train&-0.457&0.137&-3.097&0.002**&&\\
AGE\_65 <age\_Train&-0.143&0.113&-0.961&0.336&&\\
AGE\_age<=24\_Train&0.506&0.147&3.667&0.000**&&\\
AGE\_39<age<=54\_SM&-0.422&0.123&-3.938&0.000**&&\\
AGE\_54<age<= 65\_SM&0.230&0.089&1.874&0.061*&&\\
AGE\_65 <age\_SM&0.071&0.068&0.101&0.920&&\\
AGE\_age<=24\_SM&-0.247&0.095&-3.257&0.001**&&\\
INCOME\_over 100\_Train&0.281&0.175&2.589&0.010**&&\\
INCOME\_under 50\_Train&-0.137&0.131&0.269&0.788&&\\
INCOME\_unknown\_Train&-0.177&0.155&-0.030&0.976&&\\
INCOME\_over 100\_SM&0.014&0.132&1.141&0.254&&\\
INCOME\_under 50\_SM&0.070&0.093&2.210&0.027**&&\\
INCOME\_unknown\_SM&0.120&0.112&2.301&0.021**&&\\
\bottomrule
\end{tabular}\caption{Multinomial Logit Model Regression Results for PCA model}\label{pca_coeffs}\thispagestyle{empty}

\end{table}
\end{center}

\begin{table}[!htbp]
\scriptsize
\begin{tabular}{lc|ccc|HccHH}
&&\multicolumn{3}{c}{Training set} &\multicolumn{3}{c}{Test set}\\
&\# &Loglik.&$R^2$&$\bar{R}^2$&AIC&Loglik.&$R^2$&Robust $R^2$&AIC\\
Model&param.& (LL0=-6,642.8)&&&&&&&\\
\midrule
\midrule
Original &14&-4,695.8&0.284&0.282&9,419.6 &-1,486.2 &0.279&0.274&3,000.5 \\
Dummies&232&\textbf{-3,956.9} &\textbf{0.397}&\textbf{0.356}&\textbf{8,377.8}&-3,691.0 &-0.789&-1.02&7,846.2 \\
Dummies reduced&42&-4,386.9 &0.331&0.325&8,857.8&-1,429.3 &0.307&\textbf{0.292}&\textbf{2,942.5} \\
PCA&39&-4,407.9&0.328&0.322&8,893.8&-1,427.0&0.308&0.294&2,931.9\\
\midrule
Embeddings (best)&39&-4,397.0&0.330&0.324&8,871.9&\textbf{-1,389.9}&\textbf{0.326}&\textbf{0.312}&3,076.3\\
Embeddings (mean)&39&-4,397.0&0.330&0.324 &8,871.9&\textbf{-1,415.7}&\textbf{0.313}&\textbf{0.300}&\textbf{2909.5}\\
(Std. dev.)&& (10.7)&(0.001)&(0.001)& (21.4)&\textbf{(11.3)}&\textbf{ (0.005)}&\textbf{ (0.006)}&\textbf{ (22.6)}\\
%Original &14&-4,659.9&0.299&0.296&9,347.702 &-1,504.9 &0.283&0.279&3,037.8 \\
%Dummies&256&\textbf{-4,017.2} &\textbf{0.395}&\textbf{0.357}&\textbf{8,546.352}&-1,702.7 &0.189&0.072&3,917.5 \\
%Dummies reduced&42&-4,357.9 &0.344&0.338&8,799.7&-1,414.0 &0.327&\textbf{0.312}&\textbf{2,912.1} \\
%PCA&157&-4,154.3&0.375&0.351&8,622.615&--1,392.5&0.337&0.281&3,098.9\\
%\midrule
%Embeddings (best)&157&-4,131.6&0.378&0.354&8,577.1&\textbf{-1,372.1}&\textbf{0.347}&\textbf{0.292}&3,058.3\\
%Embeddings (mean)&157&-4127.5&0.379&0.355 &8,569.0&\textbf{-1,387.9}&\textbf{0.339}&\textbf{0.284}&\textbf{3,090.0}\\
%(Std. dev.)&& (17.5)&(0.003)&(0.003)& (35.0)&\textbf{(11.9)}&\textbf{ (0.005)}&\textbf{ (0.006)}&\textbf{ (23.7)}\\
\bottomrule
\end{tabular}\caption{Summary of results}\label{res_summary}
\end{table}
 %We will later
% R^2 test=0.301106 & adjusted R^2 test=0.258163
%AIC test=3854.072657 & BIC test=4685.262030
%Results of best train model=
%LL=-1784.036329, R^2=0.301106, adjr^2=0.258163, AICA3854.072657, BIC=4685.262030
 
The embeddings model seems to present the most consistent performance results across training and testing. On the training set side, considering that we have different number of parameters in different models, robust R square ($\bar{R}^2$) should be the main metric. While the dummies model still manages to have the highest value, it is clearly overfitting by looking at the test set performance. On the other hand both the ``best" embeddings model and the mean observed performance rank very close while outperforming all others in the test set. The differences are not dramatic though, and as we will see in the next section, they depend very much on the unbalance of dataset sizes for training the embeddings VS the model estimation. We also remind that the PCA model gave an acceptable balance in terms of performance in test set and significance of the coefficients.

\subsection{Embeddings for efficient survey data usage}

It is well-known that detailed travel surveys tend to become very burdensome for the respondent, and it is common to oversample in order to compensate for that problem. Often, one ends up with plenty of very incomplete records. If we designed a survey to guarantee minimal burden answers at the beginning (e.g. in Swissmetro example, ask simple quick questions like ``where are you going?", ``in which transport means?") and then entering into detailed (incentive-based?) questions, we could efficiently capture more in breadth data to learn embeddings on some of the variables, while collecting also in-depth data for a smaller set of respondents. To illustrate the impact of such an approach, we simulate the scenario where we use the entire Swissmetro dataset (minus test set) to get OD,  TICKET and mode choice information (let's call it ``light" survey). From this dataset, we also selected a sub-sample that corresponds to the ``detailed survey", with all other variables.  

We need to note that the training of embeddings is based exclusively on the ``light" survey data, therefore we cannot take advantage of our comprehensive model, that can use all data (from the ``detailed" survey) to remove their effects. In other words, our model resembles more the simpler one in Figure \ref{generalidea} than the one in Figure \ref{TravelBehaviorEmbeddings}. 

\begin{figure}[htbp]
\begin{center}
\includegraphics[width=4.8in]{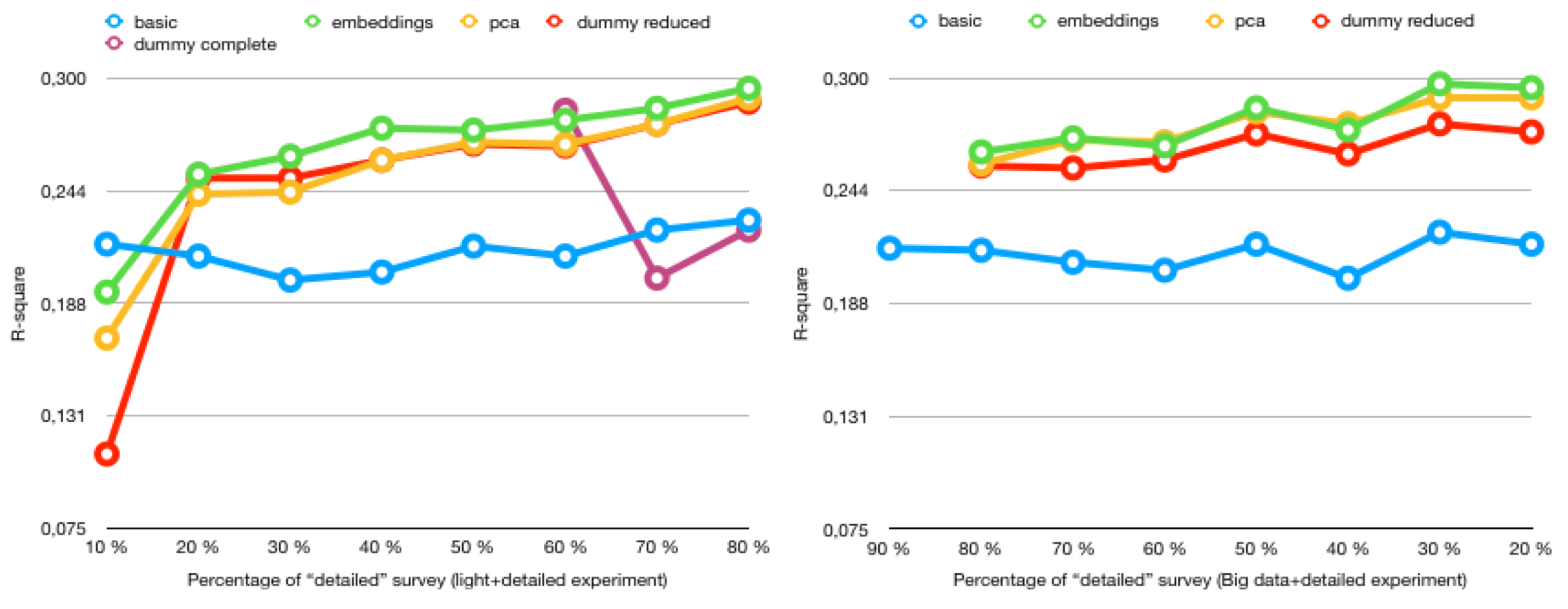}
\end{center}
\caption{R-square performance with percentage of ``expensive" survey. Left: light+detailed survey; Right: Big Data+detailed survey Note: Absence of data points means either negative R-squared, or model not possible to estimate (e.g. due to singular matrix)}\label{cheap_expensive1}
\end{figure}

Figure \ref{cheap_expensive1} (left) shows how the R-square in test set varies with the percentage of ``detailed" survey collected. Understandably, when there is very little data (10\%), the original (``basic") model performs best, but it is quickly outperformed by the embeddings model, which is practically always the best in the test. On the other hand, the PCA and simpler dummy models seem to show comparable results, especially at higher percentages. This follows our intuition, that with a sufficiently big and balanced dataset, a dummy variable model can show to be sufficient, and that PCA is also an acceptable approach. In agreement with this intuition, we can see that ``dummy complete" model (which is the one with higher dimensionality) is only possible to estimate in the higher percentages, and even then shows some erratic performance. 

%\begin{figure}[htbp]
%\begin{center}
%
%\end{center}
%\caption{Big data survey + expensive survey: R-square performance with percentage of ``expensive" survey (absence of data points means either negative R-squared, or not possible to estimate (singular matrix))}\label{big_data_expensive1}
%\end{figure}

Another hypothetical application scenario is when there is plenty of data on mobility from broad sources, like telecom or smartcard data. This often called ``big data" has the drawback of being superficial in terms of travel details, but it may already contain enough information to encode embeddings. For example, in our Swissmetro example, we could have had access to OD data from telecom and inferred modes according to Bachir et al \cite{BACHIR2019254}. While the usability of such a source could be questionable as a direct ground truth for a choice model, in the case of embeddings what we look for is to capture consistent similarity of variables (e.g. ODs) with respect to mode choice, which is ultimately what embeddings are about. Assuming that mode inference in \cite{BACHIR2019254} is at least unbiased even if with high error, the specific choice models could be properly estimated with the more detailed (smaller) dataset. Figure \ref{cheap_expensive1} (right) shows the results of simulating a scenario where we split the Swissmetro dataset into different ratios of ``big data" survey and ``detailed survey". 

Again, we note that the embeddings (and PCA) model is based only on OD+mode information, thus it may be confused by effects of other variables. Still it performs consistently better than the others, though admittedly not in a substantial way. 

\section{Discussion}

The nature of our embeddings formulation is rather simple. Indeed, one can see it just as a linear algebraic projection into a lower dimensional latent space, and if one truly aims to keep full interpretability as we know it in RUM, it is difficult to go much beyond. But as pointed out earlier, the concept of embeddings can be more flexible. One could introduce a more complex structure between the embeddings layer and the input layer. A simple way to see this is with variable interactions. For example, one can add an extra layer that receives the vectors of two input categorical variables, and then generates all possible interaction combinations, then feeding it into embeddings. Or we can just arbitrarily add non-linear layers, and select a low dimensional one to represent our input (e.g. Nguyen et al \cite{nguyen2018sqn2vec} do this for sequential data). Another apparent limitation that can be relaxed here relates to the discrete nature of the input variables. There is no reason not to incorporate a softmax layer directly after a numeric input to discretize it, and then directly apply embeddings. 

Another interesting concept to explore is seeing embeddings as ``proxies" to information that is otherwise inaccessible. Take for example a departure time model. Given a certain trip purpose, its common sense time window (we are at restaurants at certain time, at work in other times, pick up/drop off on others, etc.) will be relevant. While this may be implicit in the survey data itself, through correlations with the purpose variable, one could make a more direct association through embeddings. In this case, our embeddings algorithm would need to be changed at the output. Instead of a single softmax layer, we would have T (T=24, if we have 1 hr resolution) binary ones, thus the output variable representation consists of a vector of T ``1" and ``0'' corresponding to an activity observation, or a schedule. One could simply use Google's ``popular" times graphs for this. The interesting thing with this idea, is that one variable (Trip Purpose) would now incorporate information on its time of day activity using embeddings while at the same time reducing its dummy variable dimensionality. 

Finally, a third promising benefit of embeddings relates with its proven value in language translation. The idea is that, in the latent space, words from different languages that relate to the same concept should be close together. The way to enforce this in embeddings learning, is by having ``anchor" words, a subset of words that are well-known matches between the two languages. These would have fixed embeddings during the learning process. But what does this mean from the perspective of choice behavior? In principle, if all related models in the field are indeed consistent, one could use this idea to get different common variables (e.g. trip purpose, education level, family type) to ``live" in the same space. This means that we could directly apply and test choice models estimated from different places, compare variables from different surveys, and maybe more importantly grow our common embeddings representation together as a community. 

All of the above is quite exciting, and our model should be seen as just a first step in this direction, but we should reemphasize the limitations that we already find in our model, in light of our experiments. It is clear to us that the stochastic nature of the algorithm is not desirable. If we look at all the 300 embeddings runs as reported in Table \ref{res_summary}, we find that the test set loglikelihood performance ranged from a promising -1365.0 to a terrible -1466.0. These are obviously outliers, and our selection process (based on the development set) was able to select one clearly in the upper half of the list, but we can see the sensitiveness to initial conditions.  

It was also clear that the size of dataset, especially the proportion of training set size to number of categories to encode is determinant. If we have few categories, and a large dataset, we can directly use dummy variables. When a modeler makes a survey, she will already have this in mind, and our dataset itself was generally well-behaved except for the OD variable that was not actually explicitly used in the original paper. Maybe considering the opportunity of embeddings modelers will be more ambitious in terms of model complexity (and survey design). 

%
%. Mention previous experiences with another dataset (TU Travel survey), substantially bigger, with more impressive results here and there, but model not generally consistent. LoS variables required extra work that was not justified.
%- Clear that our model can learn something that others don't, while keeping interpretability

%\section{Beyond alternative encoding discussion}
%\subsubsection{Time signatures}
%\subsubsection{Social dimension}
%\subsubsection{Activity sequence}
%\subsubsection{Context}
%
%
%\section{Can small surveys tell us more?}
%
%\section{Wide open opportunities}
%
\section{Conclusions}
This paper introduced the concept of Travel Embeddings, that builds on earlier work from the field of Machine Learning. The general idea is to represent categorical variables using embedding encoding instead of other traditional methods, such as dummy variables or PCA encoding. Our method keeps the interpretability of the traditional approaches. We showed that there are good benefits in terms of model performance both at training and test set sides, generally outperforming the other models tested. There are, however, limitations to consider, namely the stochastic nature of the algorithm that required a careful experimental design. 

We see this paper as the starting one on the use of embeddings in behavior choice modeling, and provide a new Python package for others to use. This is quite a flexible framework to collaborate on, and thus we encourage others to extend it, perhaps towards a community with golden standard shared resources, as already happens with the Natural Language and Image Processing communities, to give an example. 

%\section*{References}

\bibliography{mybibfile}

\end{document}